%% file: main.tex
\documentclass[conference]{IEEEtran}
\input{src/header.tex}

\input{src/frontmatter.tex}

\begin{document}
    \maketitle
    \pagestyle{plain}

    \input{src/abstract.tex}

    \section{Introduction}
    \label{sec1}
    \input{src/section/1/section-1.tex}

    \section{Background on IPFS}
    \label{sec2}
    \input{src/section/2/section-2.tex}

    \section{Related Work}
    \label{sec3}
    \input{src/section/3/section-3.tex}

    \section{Active Sybil Attack on IPFS}
    \label{sec4}
    \input{src/section/4/section-4.tex}

    \section{Sybil-Resistant DHT Store (SR-DHT-Store)}
    \label{sec5}
    \input{src/section/5/section-5.tex}

    \section{Enhanced Mitigation Strategy}
    \label{sec6}
    \input{src/section/6/section-6.tex}

    \section{Conclusion}
    \label{sec7}
    \input{src/conclusion.tex}

    \section{Ethics Consideration and Responsible Disclosure}
    \label{ec}
    \input{src/ethics-consideration.tex}

    \section*{Acknowledgement}
    \label{ack}
    \input{src/acknowledgment.tex}

    \bibliographystyle{IEEEtran}
    \bibliography{bibliography.bib}

\end{document}

%% file: src/frontmatter.tex
\title{Active Sybil attack and efficient defense strategy \\ in IPFS DHT}

\author{
    \IEEEauthorblockN{
        Victor Henrique de Moura Netto,
        Thibault Cholez and
        Claudia-Lavinia Ignat
        }
    \IEEEauthorblockA{\textit{Université de Lorraine, CNRS, Inria, LORIA, Nancy, 54000, France}\\
    \texttt{\{\href{mailto:victor-henrique.de-moura-netto@inria.fr}{victor-henrique.de-moura-netto}, \href{mailto:thibault.cholez@inria.fr}{thibault.cholez}, \href{claudia.ignat}{claudia.ignat}\}@inria.fr} \\
    }
}

%% file: src/abstract.tex
\begin{abstract}
    The InterPlanetary File System (IPFS) is a decentralized peer-to-peer (P2P) storage built on Kademlia, a Distributed Hash Table (DHT) structure commonly used in P2P systems and known for its proved scalability. However, DHTs susceptible to Sybil attacks, where a single entity controls multiple malicious nodes. Recent studies have shown that IPFS is affected by a passive content eclipse attack, leveraging Sybils, in which adversarial nodes hide received indexed information from other peers, making the content appear unavailable. Fortunately, the latest mitigation strategy coupling an attack detection based on statistical tests and a wider publication strategy upon detection was able to circumvent it.

    In this work, we present a new active attack in which malicious nodes return semantically correct but intentionally false data. The attack leverages strategic Sybil placement to evade detection and exploits an early termination in the actual Kubo, the main IPFS implementation. It achieves to fully eclipse content on recent Kubo versions. When evaluated against the most recent known mitigation, it successfully denies access to the target content in approximately 80\% of lookup attempts.

    To address this vulnerability, we propose a new mitigation called SR-DHT-Store, which enables efficient, Sybil-resistant content publication without relying on attack detection. Instead, it uses systematic and precise use of region-based queries based on a dynamically computed XOR distance to the target ID. SR-DHT-Store can be combined with other defense mechanisms, fully mitigating passive and active Sybil attacks at a lower overhead while supporting an incremental deployment.
\end{abstract}

%% file: src/section/1/section-1.tex

Since the early 2000s, peer-to-peer (P2P) networks have been known to be vulnerable to Sybil Attacks \citep{douceur-sybil-2002}, in which an attacker instantiates many fake peers under their control to disrupt the network's operations. Structured P2P networks based on Distributed Hash Tables (DHTs), such as Kademlia \citep{maymounkov-kademlia-2007}, are particularly threatened as each peer is responsible for indexing a range of content space according to its position in the DHT. This vulnerability leads to localized attacks, where an attacker forges IDs to insert multiple Sybil nodes into a specific region of the DHT, gaining control over the indexing mechanism and denying access to targeted content \citep{steiner-exploiting-kad-possible-2007, cholez-monitoring-controlling-content-2010}. To counter such threats, several defense mechanisms have been proposed \citep{cholez-detection-mitigation-localized-2013, sridhar-content-censorship-2023}, typically relying on evaluating the distribution of the peers around an ID using statistical tests of relative entropy, to detect anomalies indicative of Sybil activity. Additionally, they may modify the lookup strategy to either avoid suspicious nodes or reduce their influence by diluting their weight in the content search process.

In recent years, the InterPlanetary File System (IPFS) \citep{benet-ipfs-2014} has emerged as an attempt to build the storage layer for the decentralized World Wide Web (WWW). IPFS is a suite of protocols designed for data exchange, built on the principles of content addressing and peer-to-peer networking. Data is never directly uploaded to the network, only references to the providers of the content, who are contacted directly for data exchange. The primary implementation of IPFS is Kubo \citep{ipfs-kubo} (formerly known as go-ipfs), written in Go. IPFS is used in a variety of applications, including Berty \citep{berty}, a peer-to-peer messaging app that operates without central servers; DTube \citep{dtube}, a decentralized video-sharing platform; and Filecoin \citep{filecoin}, a cryptocurrency-based decentralized file storage network. The core networking functionality of IPFS has been extracted into a separate project called libp2p \citep{libp2p-website}, an open-source, modular framework for building P2P applications.

Recent studies \citep{sridhar-content-censorship-2023, cholez-sybil-attack-strikes-2024} have shown that the main IPFS client has so far neglected to implement security mechanisms against Sybil attacks, and was consequently highly affected by a basic Sybil attack scenario. In such attack, an attacker can eclipse content from the network by precisely inserting Sybil nodes with forged IDs. To address this issue, the authors of \citep{sridhar-content-censorship-2023} proposed a detection and mitigation strategy, which reduces Sybil impact and counters censorship attempts. Their approach involves extending the set of peers that receive publication records by performing region-based queries, based on their common prefix length (CPL) with the CID. With the increasing adoption of IPFS in both practice and research works \citep{yang-blockchain-based-file-replication-2024, chen-tackling-data-mining-2025}, such a Sybil attack may disrupt any system that relies on the protocol.

In this paper, we propose a new attacker model that optimizes both the number and position of Sybil nodes while remaining below the detection threshold of the proposed defenses. As a result, we show that the majority of publication records can still be intercepted by them without triggering the defense mechanism. We exploit this vulnerability by introducing active behavior in Sybil nodes, which now advertise fake records to trigger early termination in IPFS's lookup process, leading to an attack success rate of 80\%. In response, we propose a new Sybil-resistant strategy for storing data in the DHT. This approach relies on an efficient way for calculating the XOR distance where the peers responsible for a record should be located, combined with a systematic use of region-based queries that we have optimized to significantly reduce overhead. Our mechanism called SR-DHT-Store, when used in combination with other mechanisms such as disjoint lookup paths \citep{baumgart-kademlia-practicable-approach-2007}, fully mitigates the active attack and defines a new best practice for securely publishing data in a DHT. For the sake of reproducibility and to benefit to the community, all source code related to the attack, proposed mitigation and evaluation is publicly released as open source.\footnote{\url{https://gitlab.inria.fr/loreley-team/active-sybil-attack-and-efficient-defense-strategy-in-ipfs-dht}.}

To summarize, our main contributions are the following:
\begin{itemize}
    \item A new Sybil Attack strategy in IPFS DHT featuring:
    \begin{itemize}
        \item a combinatorial optimization algorithm for inserting Sybil nodes while remaining undetected against the latest defense mechanisms proposed by Cholez et al. \citep{cholez-detection-mitigation-localized-2013} and Sridhar et al. \citep{sridhar-content-censorship-2023};
        \item an active attacker model that exploits a vulnerability to prematurely abort the content lookup process, implemented and evaluated on the real IPFS network;
    \end{itemize}
    \item A new Defense strategy against Sybil attacks in DHTs featuring:
    \begin{itemize}
        \item an efficient method for estimating the appropriate distance to peers responsible for content, based on the XOR metric rather than the Common Prefix Length (CPL);
        \item a systematic use of region-based queries with reduced overhead for content advertisement;
        \item an evaluation of combined defense mechanisms against Sybil attacks, including our proposed mitigation strategy, conducted on the real IPFS network.
    \end{itemize}
\end{itemize}

The remainder of the paper is organized as follows. Section \ref{sec2} and Section \ref{sec3} respectively present the technical background on IPFS and related work on Sybil attacks and defense mechanisms for DHTs, with a specific focus within the context of IPFS. Section \ref{sec4} presents and evaluates our active Sybil attack scenario and its results. Our Sybil-resistant publication strategy for the DHT, SR-DHT-Store, is described and evaluated in Section \ref{sec5}. In Section \ref{sec6}, we further enhance DHT resilience by combining our proposed publication strategy with additional mechanisms, in particular S/Kademlia, before concluding the paper in Section \ref{sec7}. Finally, in Section \ref{ec}, we discuss the ethical considerations of our work.

%% file: src/section/2/section-2.tex
In this section, we provide a comprehensive overview of the key concepts necessary to understand this work. We begin by introducing peer and content identifiers, followed by the fundamental principles of the DHT. As a core component of IPFS, we then examine the DHT’s internal structures, such as the routing table (RT), along with the mechanisms used for content addressing and retrieval.

\subsection{Overview}

IPFS uses a distributed content indexing mechanism built on a DHT to make this discovery efficient, where peers can be quickly located by any node searching for the content. Actual data is never uploaded to the network, instead, references to the data are distributed to strategically positioned peers. These references are known as provider records (PRs), and they map a content identifier (CID) to a peer identifier (PID), enabling the requesting peer to locate and contact the appropriate provider for content retrieval.

When sharing content, the publishing node communicates the content identifier (CID) through an out-of-band channel. A peer wishing to download the file then searches the network to locate the corresponding record associated with the CID and directly contacts the listed providers for the actual data exchange. If none of the providers associated with the content are reachable, the file becomes temporarily inaccessible. To ensure persistent availability, pinning services such as Piñata \citep{pinata} and Filebase \citep{filebase} can be used. These services act as highly reliable providers with minimal downtime.

\subsection{Identifiers}

Each piece of content in the network is assigned a unique identifier, known as a content identifier, derived from the hash of the file’s content. This hash is computed from a Merkle Directed Acyclic Graph (MerkleDAG) \citep{merkledag-ipfs-docs}, which splits the content into fixed-size segments, called chunks, and organizes them into a tree-like structure. The root of this tree, is obtained from the hash of all the chunks and serves as the CID for the content. As a result, any modification to the content produces an entirely different CID, allowing recipients to verify the authenticity and integrity of the data by recalculating the hash of the received content. The root hash is encoded using the Multiformats standard \citep{multiformats}, producing a self-describing, future-proof value that supports multiple hashing algorithms, encoding schemes, and serialization formats.

Peers are identified by peer identifiers, which are derived from the hash of a locally generated public key. This PID is formatted similarly to CIDs, using the Multiformats standard. To establish a secure connection, peers exchange self-signed certificates that include their public key, along with a signature generated using their respective private key, as an extension to the certificate. This ensures the authentication of both the source and destination peers while establishing a secure connection for data exchange.

\subsection{DHT}

Nodes and content become accessible once they are inserted into the DHT, which operates within a 256-bit key space. This distributed hash table contains pairs of keys and values, distributed along multiple nodes in the network. The keys are the SHA-2 hash of a CID or PID, ensuring a fixed-size, comparable value regardless of their original format. There are two main elements inserted into this DHT: provider records (PR) and peer records.

When providing content to the network, a \textbf{provider record}, containing the pair \{\textit{key:} provided CID, \textit{value:} provider PID\}, is distributed to selected peers. These peers store this record in their local hash tables, which are periodically cleaned to avoid retaining references to unused content due to network churn. By default, to ensure the continuous availability of the content, the provider must periodically announce the record, a process known as content pinning \citep{content-pinning-ipfs-docs}. 

After identifying the content provider from an obtained PR, the node must retrieve the provider’s IP address to initiate the content exchange. This is done by searching for the corresponding \textbf{peer record}, which maps \{\textit{key:} PID, \textit{value:} multiaddresses\}. The values of those pairs are multiaddresses \citep{multiaddr-multiformats}, a standardized format used to represent a peer’s contact information, including all supported protocols, IP addresses, and ports.

To search and provide records in the DHT, nodes perform a \textbf{DHT lookup}, an iterative process to find the $k$ closest nodes to a target identifier. Initially, the $\alpha$ closest nodes to the target on its routing table are queried in parallel for the closest nodes they know to the identifier. All responses are stored in a shared data structure, ordered by distance, and used in subsequent lookup iterations. The process continues until the closest $\beta$ nodes found have already been queried and have successfully responded. IPFS follows the Kademlia protocol specifications \citep{maymounkov-kademlia-2007}, using the XOR metric to measure distance between identifiers, and the parameters: $k = 20$, $\alpha = 10$, and $\beta = 3$ \citep{amino-config-libp2p}.

\subsection{Routing Table}
\label{sec2:ssec:routing-table}
When joining the public network, a bootstrap process begins by contacting one of the official IPFS bootstrap nodes. They serve as entry points for populating the node's routing table and announcing their presence to the network. The RT consists of $i$-buckets, where $i \in \{0, 255\}$, each capable of storing up to $k$ nodes that share exactly $i$ leading common bits with the node’s own identifier. In this paper, $i$ is also referred to as the common prefix length (CPL) relative to another identifier. 

To keep all buckets updated, the routing table is refreshed every ten minutes, as well as immediately after bootstrap. During each refresh, the node removes peers that are unreachable, offline, or have not demonstrated usefulness within a specified time window \citep{dht-ipfs-docs}. Each bucket $i \in \{0, 255\}$ should be updated by performing a lookup toward a randomly generated identifier that shares exactly $i$ common prefix bits with the node's PID. In practice in IPFS, only the first buckets $0 \leq i \leq 15$ are actively refreshed in this manner. Peers discovered during the process are organized into the appropriate buckets. This procedure is followed by a final lookup toward the node's own identifier to refresh the remaining buckets with $i \geq 16$.

In addition to the RT, libp2p nodes maintain a structure known as the Swarm, which consists of a collection of interconnected peers. These peers function similarly to a cache. When searching for information, the node first queries its Swarm, which may already possess the requested data, eliminating the need to perform a DHT lookup to find a provider.

\subsection{Content Addressing and Retrieval}
\label{sec2:subsec:content-addressing-and-retrieval}

\begin{figure}[ht]
    \centering
    \includegraphics[width=\columnwidth]{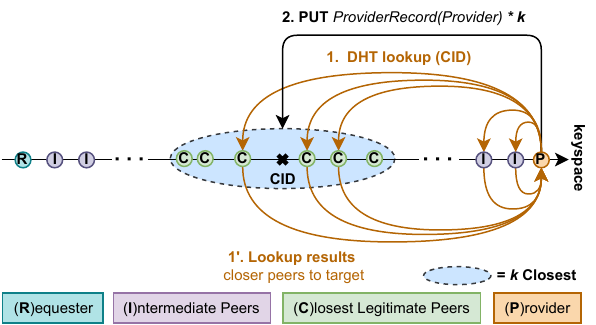}
    \caption{Content provide in IPFS DHT. This example assumes $k = 6$.}
    \label{sec2:fig:put-ipfs}
\end{figure}

To make content available on the network, provider records are distributed to the $k$ closest nodes to the precomputed CID. These nodes are obtained through a DHT lookup, and each of them is contacted individually to receive the record, becoming \textit{resolvers} for the content. This process is illustrated in Figure~\ref{sec2:fig:put-ipfs}.

\begin{figure}[ht]
    \centering
    \includegraphics[width=\columnwidth]{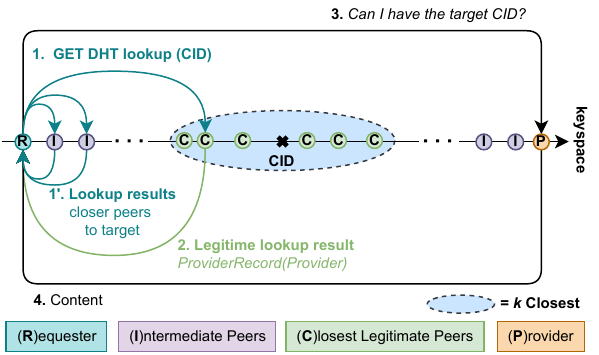}
    \caption{Content retrieval in IPFS DHT. This example assumes $k = 6$.}
    \label{sec2:fig:get-ipfs}
\end{figure}

The retrieval lookup is slightly modified compared to a standard Kademlia lookup: it opportunistically requests both the provider record of the content and the $k$ closest peers known to each queried node. The lookup continues until a valid provider is found, and the content is retrieved, or until ten provider records have been collected, a termination further exploited in this paper to perform an active attack. All data exchanges in IPFS are made using Bitswap \citep{bitswap-ipfs-docs}. This retrieval process is illustrated in Figure \ref{sec2:fig:get-ipfs}.

%% file: src/section/3/section-3.tex
In open and fully distributed P2P networks such as IPFS, there is no centralized authority, no privileged peers, and no mechanism to verify the association between a peer identifier and a specific user or device. This lack of centralization enables a single entity to assume multiple malicious identities, known as \textbf{Sybil} nodes \citep{douceur-sybil-2002}. By strategically positioning these nodes, an attacker can gain control over the distributed network and carry out either passive or active attacks, leading to content or nodes being eclipsed from the network. In this section, we review past Sybil-based attacks and corresponding countermeasures applied to Kademlia DHT and IPFS.

\subsection{DHT Attacks}

Eclipse attacks, named by \citet{Singh et al.}{singh-eclipse-attacks-overlay-2006}, aim to isolate a target node or content from the network by surrounding it with malicious peers, effectively hiding it from the rest of the system. In this section, we focus on studies that explore two main strategies for executing this attack by either manipulating the routing table or disrupting internal DHT operations.

Initially focusing on routing table manipulation, \citet{Wang et al.}{wang-attacking-kad-network-2008} implemented a targeted routing attack against the Kad network, also based on the Kademlia DHT. They introduced a method for hijacking existing connections in a peer’s routing table by impersonating legitimate nodes. By querying victims and mapping their routing tables, the attacker sends spoofed messages to update the contact information of the nodes listed in the RT. As a result, the victim becomes dependent on malicious nodes to communicate with the rest of the network. This attack exploits Kad's lack of identifier verification during peer exchanges, an issue addressed in IPFS.

Another variation of the eclipse attack exploits the interval structure of the DHT. \citet{Steiner et al.}{steiner-exploiting-kad-possible-2007} describe an attack in which adversaries eclipse content by inserting malicious nodes near the target identifier using the XOR metric. Since these malicious nodes become the only resolvers for the content, they can control how they respond to queries, effectively dictating access to the targeted data. \citet{Cholez et al.}{cholez-monitoring-controlling-content-2010} demonstrated content retrieval can be controlled using only $k$ Sybil nodes by efficiently eclipsing each of the $k$ file records sent during a publication. \citet{Wang et al.}{wang-attacking-kad-network-2008} proposed a more active strategy, where attackers respond to lookup requests with an excessive number of bogus results. This triggers early termination due to a result limit, preventing legitimate queries from completing. In the work of \citet{Kohnen et al.}{kohnen-conducting-optimizing-eclipse-2009}, attackers respond to lookup requests with Sybil identities that are only slightly closer to the target than the current best-known nodes. This causes the lookup process to make minimal progress and continue indefinitely until it times out, at which point the content is considered unreachable.

\subsection{IPFS Attacks}

In this subsection, we provide an overview of documented attacks targeting the IPFS network. We should note that attacks originally designed for Kademlia are not necessarily applicable to IPFS, as IPFS includes additional security mechanisms that were not present in the original Kademlia design.

\textbf{Network Partitioning.} The first documented attack on IPFS was presented by \citet{Prünster et al.}{prunster-total-eclipse-heart-2022}. The authors described a method to eclipse a victim node by fully populating its routing table with Sybil nodes, making it dependent on malicious peers to contact the rest of the network. During DHT lookups, the victim sends queries to the closest nodes it knows from to the target identifier, which in this scenario are all malicious, allowing the attacker to drop or filter requests. Once eclipsed, the victim's only remaining connections to the legitimate network are through its peer Swarm. 

They evaluated two strategies: a naïve approach and a more effective one. In the naïve approach, the attackers continuously ping the victim’s first $\log_2(n)$ buckets until all reliable entries in the victim's routing table are dropped and replaced with Sybil nodes. In the more effective technique, Sybil nodes gain priority over honest peers by sending unsolicited data to inflate their usefulness scores. As a result, the victim’s connection manager evicts legitimate peers in favor of the seemingly more useful malicious nodes.

Although this peer eclipse attack proved effective, it was mitigated in the go-ipfs (formerly Kubo) 0.7 update \citep{hardening-IPFS-public-2020}, as further explained in Section \ref{sec3:sec:kubo-constraints}.

\begin{figure}[ht]
    \centering
    \includegraphics[width=\columnwidth]{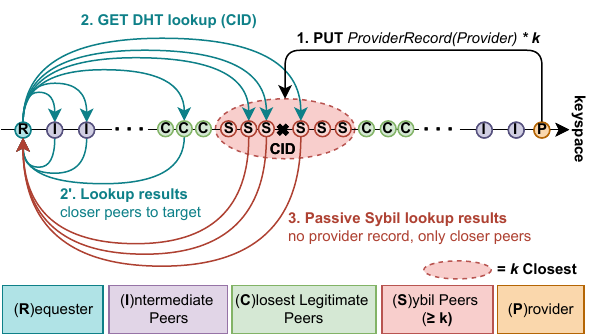}
    \caption{Content eclipse attack in IPFS DHT. This example assumes $k = 6$.
    }
    \label{sec3:fig:get-put-content-eclipse}
\end{figure}

\textbf{Content Eclipse.} The content eclipse attack was demonstrated by \citet{Sridhar et al.}{sridhar-content-censorship-2023} and \citet{Cholez and Ignat}{cholez-sybil-attack-strikes-2024}. The attacker begins by brute-forcing at least $k$ peer identifiers to generate and instantiate nodes that are closer than any legitimate peer to the target content. During the next republication interval, the provider node unknowingly sends the provider record exclusively to these malicious nodes, which are now the $k$ closest peers to the content. When the Sybil is queried during a DHT lookup, it responds with empty results, claiming no knowledge of the received records. Once the expiration time is reached for the reliable nodes that had previously stored the records, before the Sybil nodes were instantiated, the content becomes eclipsed. This attack is illustrated in Figure \ref{sec3:fig:get-put-content-eclipse}. For simplicity, the DHT lookup performed by the provider to locate the $k$ closest nodes has been omitted from the figure.

In Kubo, provider records are republished at an interval of $t_r = 22$ hours and expire after a duration of $t_e = 48$ hours. While generating appropriate PeerIDs is computationally intensive due to the large 256-bit keyspace, both studies demonstrated that it is computationally feasible. \citet{Cholez and Ignat}{cholez-sybil-attack-strikes-2024} showed that generating 20 Sybil PeerIDs sufficiently close to any target CID would take less than 1.5 hours on a standard 8-core desktop machine.

Table \ref{sec3:tab:attack-comparison} presents a summary of previous attacks in IPFS compared to our approach. As it will be shown in Section \ref{sec4}, by adopting an active behavior targeting the content indexing, we are able to launch a successful attack involving less Sybil nodes and bypassing current mitigation solutions.

{
    \definecolor{description}{HTML}{FFD966}
    \definecolor{our-approach}{HTML}{FFF2CC}
    \tiny

    \begin{table}[ht]
        \centering
        \caption{Differences between previous attacks on IPFS and our proposed approach.}
        \label{sec3:tab:attack-comparison}
        \begin{tblr}{width=\columnwidth, colspec={|X[0.2,c]|X[0.25,c]|X[0.21,c]|X[0.18,c]|X[0.18,c]|}}
            \hline
            \SetCell[r=2]{c, description} Paper & \SetCell[r=2]{c, description} Target & \SetCell[c=2]{c, description} Sybil nodes && \SetCell[r=2]{c, description} \makecell{Existing \\ Mitigation} \\
            \hline
            && \SetCell{c, description} Amount & \SetCell{c, description} Behavior & \\
            \hline
            \SetCell{c} \makecell{Prünster \\ et al. \citep{prunster-total-eclipse-heart-2022}} & \SetCell{c} \makecell{Routing Table} & \SetCell{c} \makecell{$k \times \log_2(N)$} & \SetCell{c} Active & \SetCell{c} Yes \\
            \hline
            \SetCell{c} \makecell{Sridhar \\ et al. \citep{sridhar-content-censorship-2023}} & \SetCell{c} Indexing & \SetCell{c} $45$ & \SetCell{c} Passive & \SetCell{c} Yes \\
            \hline
            \SetCell{c} \makecell{Cholez and \\ Ignat \citep{cholez-sybil-attack-strikes-2024}} & \SetCell{c} Indexing & \SetCell{c} $k = 20$ & \SetCell{c} Passive & \SetCell{c} Yes \\
            \hline
            \SetCell{c, our-approach} Our paper & \SetCell{c, our-approach} Indexing & \SetCell{c, our-approach} $< k = 20$ & \SetCell{c, our-approach} Active & \SetCell{c, our-approach} No \\
            \hline
        \end{tblr}
    \end{table}
}

\subsection{IPFS Defense Mechanisms}
\label{sec3:subsec:ipfs-defense-mechanisms}

Over the years, various strategies have been proposed to mitigate Sybil attacks. \citet{Douceur}{douceur-sybil-2002} argues that no distributed system can be fully Sybil-resistant without a centralized certificate authority, an incompatible approach with decentralized, open-source protocol like IPFS. For this reason, mitigation strategies are essential to limit the Sybil impact. This subsection reviews previously proposed detection and mitigation methods against Sybil attacks, concluding with an overview of the defenses implemented in Kubo.

\textbf{S/Kademlia.} According to the original IPFS paper \citep{benet-ipfs-2014}, the protocol should implement two key recommendations from S/Kademlia \citep{baumgart-kademlia-practicable-approach-2007} to improve DHT robustness: identity generation derived from key pairs and disjoint lookup paths. The first ensures that each node obtains its identifier from a local key pair, which is used for identity and message authentication. The second proposes that lookups must be performed along separate, non-overlapping paths, storing responses in isolated queues to reduce the impact of adversarial peers during routing.

However, as presented by \citet{Cholez and Ignat}{cholez-sybil-attack-strikes-2024}, Kubo does not enforce the second one, the disjoint lookup paths. While it performs parallel queries during lookups, all discovered nodes and provider records are aggregated into a single structure, making the process more vulnerable to Sybil attacks. This limitation is further analyzed in our work.

\textbf{Attack Detection using the K-L Divergence.} \citet{Cholez et al.}{cholez-detection-mitigation-localized-2013} propose a Sybil attack detection mechanism based on analyzing the distribution of the $k$ closest nodes to a given target. In a reliable network, nodes are expected to be randomly positioned across the DHT space, resulting in a uniform distribution of identifiers. By estimating the total number of nodes in the network, it is possible to compute the expected model distribution of the $k$ closest nodes to a given identifier. When malicious nodes are strategically positioned to occupy all $k$ closest positions, the observed distribution deviates from the expected random distribution. To quantify this deviation, the authors use the Kullback-Leibler (K-L) divergence, which measures the discrepancy between two probability distributions: the expected model distribution and the observed real distribution. The minimum value of the K-L divergence is $D_{KL} = 0$, which indicates no difference between the distributions, increasing as the discrepancy between the distributions grows.

\citet{Sridhar et al.}{sridhar-content-censorship-2023} implemented this attack detection method on IPFS. Given a network size $N$, they compute their model distribution by estimating the probability that a randomly selected closest peer $j$ has a CPL exactly equal to $x$ bits. This probability distribution, $p(x)$, is then compared with the empirical distribution $q(x)$, which is obtained from the actual $k$ closest nodes to the target content. Both distributions are compared using the K-L divergence equation, as follows:
\begin{equation}
    D_{KL} \triangleq \sum_{x \in X} q(x) \ln{\left(\frac{q(x)}{p(x)}\right)}
    \label{sec3:eq:kl-divergence}
\end{equation}

Through experiments on the IPFS network, they established a threshold value of 0.94. If the discrepancy between the observed and expected distributions exceeds this threshold, the distribution is classified as an attack. The authors prioritized minimizing false negatives to increase the likelihood of detecting an attack when defining this constant, at the cost of introducing some overhead. Their threshold is calibrated for a false positive rate of $4.4\%$ and a false negative rate of $0.81\%$.

\begin{figure}[ht]
    \centering
    \includegraphics[width=\columnwidth]{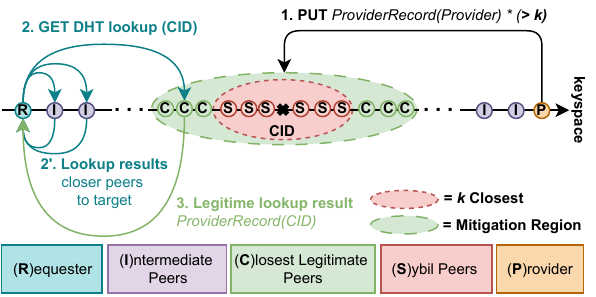}
    \caption{Region-based query mitigation strategy. This example assumes $k = 6$.}
    \label{sec3:fig:get-put-region-based}
\end{figure}

\textbf{Region-Based Queries.} \citet{Sridhar et al.}{sridhar-content-censorship-2023} additionally proposed a mitigation strategy to be applied upon attack detection. Their approach is based on the assumption that the $k$ closest reliable nodes to a content cannot be removed from the network by an adversary. To reach these reliable nodes, they introduce the concept of region-based queries. Using the previously estimated network size from the attack detection phase, they estimate $minCPL = \lceil \log_2(\frac{N}{k})\rceil$, which represents the CPL of the $k$-th closest reliable node in a truly random distribution of identifiers within the network's DHT space. They perform then lookups until all nodes sharing at least this $minCPL$ with the target are found, afterward sending PRs to each of them. As a result, regardless of the number of Sybil nodes deployed, approximately $k = 20$ reliable peers receive the provider record, significantly increasing the likelihood of a requester successfully retrieving the content. This mitigation strategy is illustrated in Figure \ref{sec3:fig:get-put-region-based}. 

Their search algorithm begins with an initial DHT lookup to identify the $k$ closest nodes to the target CID. The $k$-th closest node returned from the lookup shares a given $CPL$ with the target, meaning that all other returned nodes share the same or a greater $CPL$ with the target CID. In the subsequent iteration, a random identifier with $CPL + 1$ with the target is generated, to locate nodes farther from the target through additional lookups. This process is repeated until the $CPL$ of the $k$-th closest node is less than $minCPL$, ensuring that all nodes within the $minCPL$ region have been successfully identified.

At the time of writing, this mitigation has not been deployed in IPFS. In the following sections of the paper, we occasionally use the term ``region-based queries'' to refer collectively to both the attack detection mechanism and this associated mitigation strategy.

\textbf{Kubo Constraints.} The go-ipfs 0.7 release \citep{hardening-IPFS-public-2020} addressed the \citet{Prünster et al.}{prunster-total-eclipse-heart-2022} attack. In this version, IPFS introduced a mechanism that ensures active connections in the routing table are pruned only during idle periods, preserving long-lived, stable peers. Additionally, a diversity filter was implemented \citep{diversity-filter-ipfs}, limiting the addition of peers from the same /16 IPv4 subnetwork to a maximum of three in the routing table, and no more than two per bucket. However, this restriction applies only to the routing table and does not affect the lookup process, which remains vulnerable to encountering multiple peers with the same IP address, as highlighted in \citep{cholez-sybil-attack-strikes-2024}.

\label{sec3:sec:kubo-constraints}

A more recent addition in version 0.34.0 of go-libp2p \citep{release-0.34.0-go-libp2p} introduces a limitation of a maximum of eight concurrent connections per /16 subnetwork. This restriction prevents a peer from maintaining an excessive number of connections within the peer Swarm, complementing the existing diversity filter, which already imposes stricter limits on entries in the routing table.

While the passive content eclipse attack is likely to be mitigated in IPFS, no study has yet evaluated the network's resilience to an active content eclipse attack. An active attack involves malicious nodes actively responding to requests with semantically correct but falsified data. The following section investigates the impact of such attack on the IPFS network, with a particular focus on defeating the latest defense mechanism proposed for mitigating Sybil-based content censorship in IPFS \citep{sridhar-content-censorship-2023}.

Table \ref{sec5:tab:mitigation-comparison} presents a summary of previous mitigation techniques in IPFS compared to our approach. As it will be shown in Section \ref{sec5}, none of the existing solutions is able to mitigate our active Sybil attack, while they also exhibit a higher overhead.

\begin{review} {
    \definecolor{description}{HTML}{FFD966}
    \definecolor{our-approach}{HTML}{FFF2CC}
    \tiny

    \begin{table}[ht]
        \centering
        \caption{Differences between previous mitigations on IPFS and our proposed approach.}
        \label{sec5:tab:mitigation-comparison}
        \begin{tblr}{width=\columnwidth, colspec={|X[0.15,c]|X[0.3,c]|X[0.15,c]|X[0.2,c]|}}
            \hline
            \SetCell{c, description} Paper & \SetCell{c, description} Mitigation & \SetCell{c, description} \makecell{Message \\ Overhead} & \SetCell{c, description} \makecell{Active Sybil\\Attack \\ Mitigation} \\
            \hline
            \SetCell{c} \makecell{Prünster \\ et al. \citep{prunster-total-eclipse-heart-2022}} & \SetCell{c} \makecell{Preserve long-lived \\ nodes + IP Limit} & \SetCell{c} \makecell{None} & \SetCell{c} No \\
            \hline
            \SetCell{c} \makecell{Baumgart \\ and Mies \\ \citep{baumgart-kademlia-practicable-approach-2007}} & \SetCell{c} \makecell{Disjoint Lookup Paths} & \SetCell{c} Moderate & \SetCell{c} No \\
            \hline
            \SetCell{c} \makecell{Sridhar \\ et al. \citep{sridhar-content-censorship-2023}} & \SetCell{c} \makecell{Region-based Queries} & \SetCell{c} Moderate & \SetCell{c} No  \\
            \hline
            \SetCell{c, our-approach} Our paper & \SetCell{c, our-approach} \makecell{SR-DHT-Store \\ + client-side} & \SetCell{c, our-approach} Low & \SetCell{c, our-approach} Yes \\
            \hline
        \end{tblr}
    \end{table}
}

\end{review}

%% file: src/section/4/section-4.tex
This section begins by introducing an active attacker model targeting the live IPFS network, which exploits a premature termination condition in the current DHT lookup process implemented by the Kubo client. The model is then adapted to bypass the attack detection and region-based query mechanisms proposed in the latest defense strategy for IPFS \citep{sridhar-content-censorship-2023}. Finally, we analyze the attack's cost and evaluate its effectiveness by conducting real-world attacks against the IPFS network. 

For better readability, all parameters used throughout the paper are summarized in Table \ref{sec3:tab:parameters}. Non-essential variables used only for intermediate explanations are omitted.

{
    \definecolor{description}{HTML}{FFD966}
    \definecolor{our-approach}{HTML}{FFF2CC}
    \tiny

    \begin{table}[ht]
        \centering
        \caption{Parameters used throughout the paper for our attack and mitigation approaches.}
        \label{sec3:tab:parameters}
        \begin{tblr}{width=\columnwidth, colspec={X[0.25,c] X[0.75,j]}, rowsep=2pt}
            \hline
            libp2p Parameters & \\
            \hline
            $N$ & total network size, number of nodes.\\
            $k$ & maximum number of nodes in a bucket and provide replication factor. In libp2p, $k = 20$.\\
            $\alpha$ &concurrency parameter, i.e. the number of peers queried in parallel during a lookup. In libp2p, $\alpha = 10$. \\
            $\beta$ & resiliency parameter, i.e. the number of closest peers that must respond before the lookup terminates. In libp2p, $\beta = 3$. \\
            $t_r, t_e$ & provide record republication and expiration times. In libp2p, $t_r = 22$\,h and $t_e = 48$\,h. \\
            \hline
            Attack Parameters & \\
            \hline
            $CPL$ & common prefix length, i.e., the number of initial bits shared by two identifiers. \\
            $D_{KL}$ & Kullback-Leibler divergence between two distributions. \\
            $ns_{average}, ns_{lowest}$ & average and lowest network sizes from our experiments, respectively, $ns_{average} = 13,239$ and $ns_{lowest} = 12,347$. \\
            $t_{Ed25519}$ & time required to generate an Ed25519 key pair using our implementation on a machine equipped with a 13th Gen Intel i7-13700H processor, respectively, $t_{Ed25519} = 1.35\,\mu s$. \\
            \hline
            Mitigation Parameters & \\
            \hline
            $d_k$ & average distance to the $k$-th closest node. \\
            $Q_{d_k}$ & number of initial queries to random nodes used to obtain the first estimate of $d_k$. \\
            $L_{d_k}$ & number of lookups performed to start refining the estimate of $d_k$. \\
            $\alpha_{sf}$ & smoothing factor of the exponentially weighted moving average, which determines the impact of previous estimations. \\
            \hline
        \end{tblr}
    \end{table}
}

\subsection{Threat Model}
\label{sec3-bis}

We consider the public IPFS network $\mathcal{N}$ as a set of nodes $\mathcal{P} = \{p_1, p_2, \dots, p_N\}$, where $N = |\mathcal{P}|$ denotes the total network size. The network is composed of honest nodes ($\mathcal{H}$) and malicious nodes ($\mathcal{M}$), with cardinalities $N_H = |\mathcal{H}|$ and $N_M = |\mathcal{M}|$, such that
\begin{equation}
    \mathcal{H} \cup \mathcal{M} = \{p_1, p_2, \dots, p_N\}, \quad \mathcal{H} \cap \mathcal{M} = \varnothing.
\end{equation}

We assume that the adversary does not control the entire network, i.e,
\begin{equation}
    0 < \frac{N_M}{N} < 1.
\end{equation}

Each peer $p_i$ and content $c_i$ is identified by a unique bitstring $\in \{0, 1\}^{256}$ and the distance between any two identifiers $A$ and $B$ is defined by the XOR metric
\begin{equation}
    d(A, B) = A \oplus B.
\end{equation}
Honest nodes in the network generate their identifiers randomly by deriving them from a locally generated public key. They run unmodified versions of Kubo and follow the default provide and lookup procedures.

\textbf{Attacker Capabilities.} Malicious nodes also follow the protocol specifications, although they run modified versions of Kubo. They do not exploit implementation bugs to crash other peers, and they do not compromise honest nodes into sending false provider records. Their objective is solely to disrupt content indexing and discovery in the Kademlia DHT, without targeting node's routing tables (e.g. by poisoning $k$-bucket entries).

All $\mathcal{M}$ peers instantiated by the adversary are hosted on the same physical machine over the same IP address. Their identifiers are not randomly generated, as the honest nodes. Let $\mathcal{N}_k(c)$ denote the $k$ closest nodes to a content identifier $c$, the adversary brute-forces identifiers in order to maximize
\begin{equation}
    \max\left|\mathcal{N}_k(c) \cap \mathcal{M}\right|,
\end{equation}
while remaining below the attack detection threshold.

Sybil nodes actively participate in the DHT by both sending and responding to queries. For a provider lookup of content $c$, let $\mathcal{P}(c)$ denote the set of provider records returned during the lookup and  $\mathcal{P}_m(c)$ the subset of records returned by malicious nodes. Malicious nodes implement an active strategy by returning false but syntactically valid provider records, such that:
\begin{equation}
    \forall p \in \mathcal{P}_m(c) , \qquad p \notin \mathcal{H} .
\end{equation}

\textbf{Attacker Limitations.} The placement of Sybil nodes is constrained by the attack detection mechanism based on the Kullback-Leibler divergence. Let $D_{KL_r}$ denote the expected distribution of the $k$ closest nodes to a randomly chosen identifier in a non-attacked network, and let $D_{KL_a}$ denote the observed distribution after inserting Sybil nodes around $c$. The adversary must ensure that the divergence between these two distributions remains below the detection threshold:
\begin{equation}
    \label{sec3:eq:kl-r-a}
    KL(D_{KL_r}, D_{KL_a}) < 0.94.
\end{equation}
Consequently, the number of Sybil nodes in $\mathcal{N}_k(c)$ is bounded by this constraint.

\textbf{Attacker Goal.} Let $\mathcal{P}_h(c) \subseteq \mathcal{H}$ denote the set of honest providers for the content identified by $c$, and let $p_r$ be a requester. A DHT lookup for $c$ initiated by $p_r$ starts at time $t_i$ and terminates at time $t_f$. Let $\mathcal{P}_t(c)$ denote the set of provider records collected up to time $t$, and $P_{\max} = 10$ be the maximum number of records accepted by the lookup. In Kubo, the requester has a timeout threshold $T_{\max} = 10\;\text{s}$, after which the content is considered unreachable. We define the event of successful retrieval as:
\begin{equation}
    \mathsf{Succ}(p_r, c) = \left\{\exists p \in \mathcal{P}_t(c) : p \in P_h(c)\right\}.
\end{equation}

In the active attack, the adversary exploits Kubo's early termination mechanism. The eclipse attack succeeds if:
\begin{equation}
    \exists t \leq t_f : \left|\mathcal{P}_t(c)\right| \geq P_{\max}, \qquad \mathcal{P}_t(c) \cap P_h(c) = \varnothing .
\end{equation}
The attacker’s goal is to ensure that Sybil nodes are contacted early in the lookup process, allowing them to quickly fill $\mathcal{P}_t(c)$ with falsified records before any honest provider is discovered, while still satisfying $t_f - t_i \leq T_{\max}$.

\subsection{Attacker Model}

Our attack exploits an early termination mechanism by responding to lookup requests with a falsified list of provider records when queried about the targeted content. During a DHT lookup for a content’s provider record, the query is designed to stop once a predefined maximum number of records is retrieved. In Kubo, this limit is set to ten records, which can be exploited by a single or multiple malicious nodes positioned along the lookup path. Those nodes can respond with semantically correct but false records, causing the lookup to terminate prematurely. In our implementation, those records contain randomly generated identifiers and IP addresses.

The active approach achieves the same success rate as a passive attack when at least $k$ malicious nodes are positioned to be the closest around the target, intercepting all $k$ provider records sent. As previously demonstrated, when $k$ Sybil nodes surround the target, the content can be completely eclipsed from the DHT. However, if an adversary node consistently manages to be the first peer contacted during the victim’s lookup process toward the target, it can externally terminate the search, effectively eclipsing the content with as few as $n \leq k$ malicious peers.

Using our attack approach, any individual, organization, or government seeking to disrupt a file hosted on the public IPFS, such as a content or an entire website, can do so effectively. This constitutes a content censorship attack, as malicious nodes within the network can render specific content inaccessible to other users.

Since no mitigation has been implemented against the passive attack in the current Kubo version 0.33.0, presenting results using the active approach on this release is uninformative, as the attack will always succeed. Instead, we implemented the active attack against the most effective mitigation proposed for IPFS to counter content eclipse: the attack detection using statistical tests coupled with the region-based queries countermeasure described in \citep{sridhar-content-censorship-2023}.

\textbf{Termination Exploit.} The lookup process, implemented in the function \texttt{findProvidersAsyncRoutine}\footnote{\label{sec4:fn:routing-go}\url{https://github.com/libp2p/go-libp2p-kad-dht/blob/v0.28.0}}, declared in the file \texttt{routing.go}, includes a call to \texttt{runLookupWithFollowup}, defined in \texttt{query.go}, which contains three distinct search termination conditions. One of these terminations is an anonymous function passed as a parameter to the variable \texttt{stopFn}, at line 351 of \texttt{routing.go}, and shown in Listing \ref{sec4:lst:stop-fn}. In this function, \texttt{psSize} represents the current number of provider records found during the lookup, while \texttt{count} denotes the maximum number of accepted records, which is set to ten in Kubo. If the number of retrieved records exceeds \texttt{count}, the lookup process is terminated. This specific termination is referred to as an external stop. Once triggered, it initiates a continuous lookup process that is repeated every minute until the request is either successfully completed or canceled.

\begin{lstlisting}[
    language=go,
    float={htb},
    caption={The function \texttt{stopFn} exploited in IPFS to achieve early termination of the DHT lookup.},
    label={sec4:lst:stop-fn}]
    func() bool {
        return !findAll && psSize() >= count
    },
\end{lstlisting}

Our malicious nodes respond to queries with records containing randomly generated identifiers and IP addresses. Since the peer cannot establish contact with these nodes using the fake addresses, it attempts to discover their potential new contact information by performing additional DHT lookups.

\begin{figure}[ht]
    \centering
    \includegraphics[width=\columnwidth]{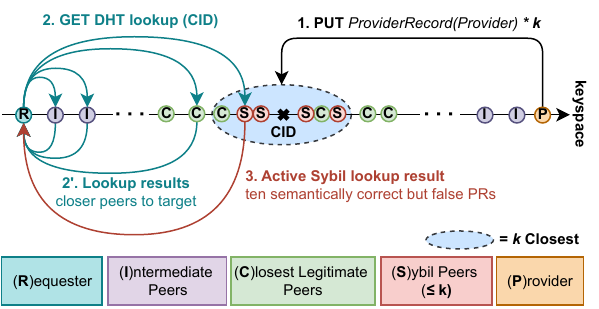}
    \caption{Active attack against the proposed mitigation. This example assumes $k = 6$.}
    \label{sec4:fig:get-put-active-attack}
\end{figure}

\textbf{Strategy.} Our strategy against the proposed mitigation  \citep{sridhar-content-censorship-2023} involves strategically inserting the maximum number of Sybil nodes while staying under the detection threshold by maintaining the appearance of a randomly generated peer distribution. The attacker begins by computing the expected model distribution, similarly to the attack detection mechanism \citep{sridhar-content-censorship-2023}, assuming an ideal scenario where all identifiers are randomly distributed across the DHT. Next, the attacker retrieves the $k$ closest nodes to the target content and calculates the K-L divergence between the observed and model distributions. We exploit the remaining margin between the divergence and the detection threshold by strategically inserting Sybil nodes without raising suspicion, taking advantage of the false negative rate. The attack strategy is illustrated in Figure \ref{sec4:fig:get-put-active-attack}.

\begin{figure}[ht]
    \centering
    \includegraphics[width=\columnwidth]{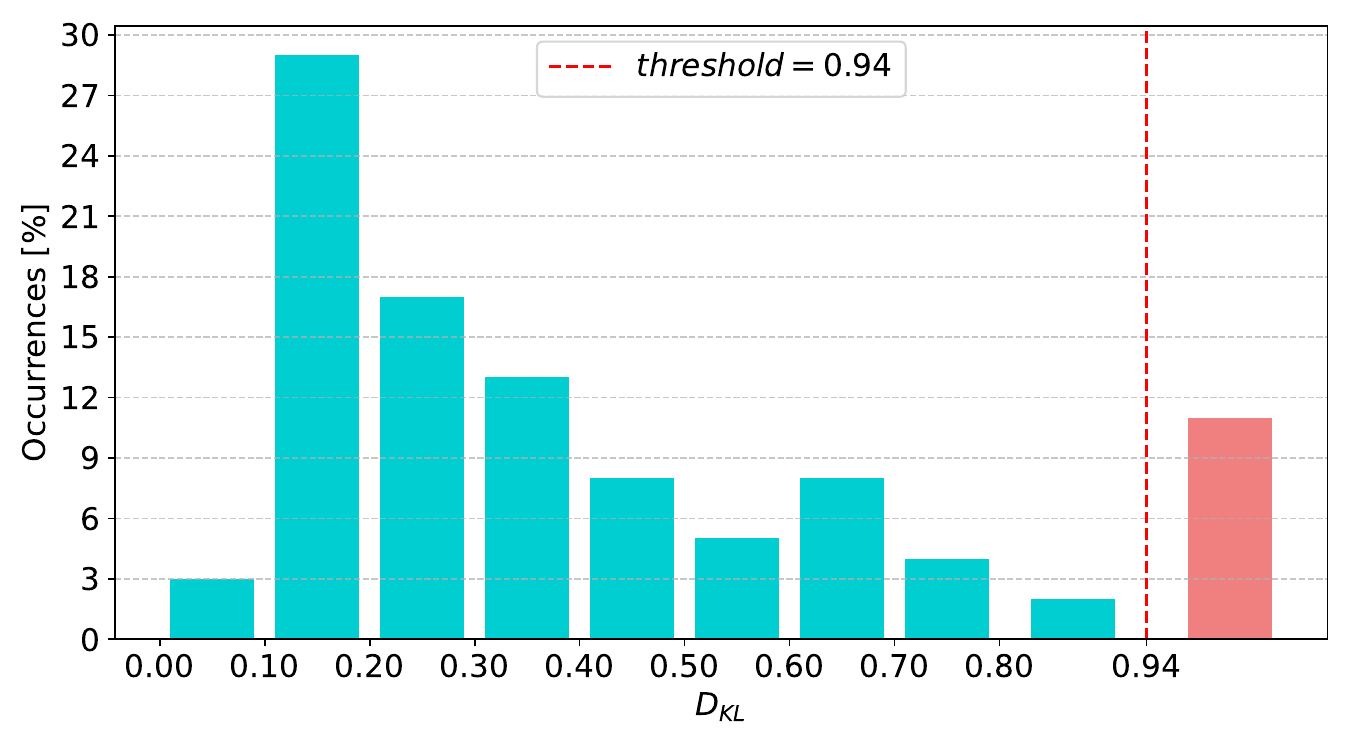}
    \caption{$D_{KL}$ distribution over 100 random DHT requests.}
    \label{sec4:fig:kl-optimization-distribution}
\end{figure}

\textbf{Average $D_{KL}$.} Around a non-attacked content, the distribution of the $k$ closest nodes in the network should resemble the model distribution, resulting in a low $D_{KL} < threshold$. To assess this, we computed the $D_{KL}$ of the $k$ closest distribution of 100 random identifiers. The model distribution was generated using the network size estimation provided by \texttt{go-libp2p-kad-dht}\footnote{\texttt{\href{https://github.com/libp2p/go-libp2p-kad-dht/blob/v0.28.0/netsize/netsize.go}{https://github.com/libp2p/go-libp2p-kad-dht/blob/v0.28.0/.../netsize.go}}}, computed prior to each test lookup. This approach avoids relying on a potentially biased network size throughout our experiments. As shown in Figure \ref{sec4:fig:kl-optimization-distribution}, the majority of $D_{KL}$ values fall below the threshold defined by \citet{Sridhar et al.}{sridhar-content-censorship-2023}, with 89\% of the samples below $D_{KL} = 0.94$. This highlights opportunities for strategically placing Sybil nodes in optimal positions while still staying below the detection threshold. On the other hand, it highlights the false positives, accounting for 11\% of the lookups.

\textbf{Optimizing Sybil Placement.} To determine the optimal positioning of Sybil nodes while minimizing the risk of attack detection, we formulate the problem as a correlated combinatorial optimization, similar to the knapsack problem \citep{pisinger-where-are-hard-kp-problems-2005}. In this problem, adding a malicious node to the distribution increases the score or profit $p_j$, while also increasing the distribution’s $D_{KL}$, which corresponds to the weight $w_j$ of the solution. Our algorithm begins by identifying the $k$ closest nodes to the target CID, counting the appearances per CPL. Using an estimated network size, we then compute the $D_{KL}$ of the current distribution.

The algorithm used is illustrated in Algorithm \ref{sec4:alg:optimize-distribution}. It contains three parameters: the results, which are empty at the start; the current nodes per CPL; and the current analyzed CPL. This last parameter starts at the CPL where instantiating at least one Sybil node would not immediately trigger attack detection due to its closeness to the target.

For each CPL in the distribution, there are $k + 1$ possible node configurations, ranging from $0$ (no nodes) to $k$ (all $k$ closest nodes). If $n_{\text{reliable}}$ reliable nodes are already present at a given $CPL$, the valid range for Sybil insertion becomes $0 \leq n_{\text{sybil}} \leq k - n_{\text{reliable}}$. Each value within the range of $n_{\text{sybil}}$ is individually tested by updating the current distribution using the \texttt{updateKClosestWithSybilNodes} function, which attempts to remove the $n_{\text{sybil}}$ farthest nodes to make space for this quantity of Sybil nodes. If successful, the function returns the updated distribution, and otherwise, it returns nothing. By evaluating the return of this function, we recalculate the $D_{KL}$ value to verify that the updated distribution still remains below the detection threshold. If the condition is met, the algorithm recursively proceeds to $CPL - 1$, until reaching $CPL = 0$, where the valid results are added to the result set for evaluation.

\begin{algorithm}[t]
    \caption{Recursive function to optimize the Sybil placement in the $k$ closest nodes from a target content.}
    \label{sec4:alg:optimize-distribution}
    \begin{algorithmic}[1]
        \Require $results$ - Stores valid distributions found
        \Require $nodesPerCpl$ - Current $k$ closest on the distribution
        \Require $currentCpl$ - Current CPL
        
        \Procedure{optimize\_sybil\_placement}{$results, \newline \hspace*{0.75em} nodesPerCpl, currentCpl$}
            \State $nodesInCpl \gets nodesPerCpl[currentCpl]$
            \For{$i = nodesInCpl$ \textbf{to} $20$}
                \State $sybils \gets i - nodesInCpl$
                \State $updatedNodesPerCpl \gets \newline \hspace*{3.75em} \text{updateKClosestWithSybils}(nodesPerCpl, sybils)$
                
                \If{$updatedNodesPerCpl$ == \textbf{null} \textbf{or} \newline \hspace*{3.75em}\text{isKLOverThreshold}(updatedNodesPerCpl)}
                    \State \textbf{break} \Comment{Invalid distribution.}
                \EndIf

                \If{$currentCpl - 1 < 0$} \Comment{Termination.}
                    \State $results.add(updatedNodesPerCpl)$
                    \State \textbf{continue}
                \EndIf
                
                \State \Call{optimize\_sybil\_placement}{$results, \newline \hspace*{3.75em} updatedNodesPerCpl, currentCpl - 1$}
            \EndFor
        \EndProcedure
    \end{algorithmic}
\end{algorithm}

In the attack detection mechanism proposed by \citet{Sridhar et al.}{sridhar-content-censorship-2023}, the comparison relies on the number of nodes per CPL between the expected model and the observed distribution, ignoring the actual proximity of nodes within each CPL based on XOR distance. To exploit this limitation, we introduce a finer level of granularity in our attack strategy by ensuring that optimized Sybil nodes are always positioned closer to the target than the honest nodes within the same CPL. This enables, for example, the replacement of all nodes located at the CPL of the $k$-th closest node by nodes having the same CPL but a smaller XOR distance to the target. As a result, the most distant nodes among the $k$ closest, sharing the same prefix, are removed from the $k$-closest set, while preserving the overall CPL distribution and thus maintaining the same $D_{KL}$.

\textbf{Score.} Based on our observations, each valid distribution should be evaluated according to the following three objectives to determine its suitability as an optimal attack distribution:
\textit{(i)} maximize the number of Sybil nodes;  
\textit{(ii)} position the Sybil nodes as close as possible to the target; and  
\textit{(iii)} ensure that the closest node to the target is a Sybil.  
These objectives lead to our scoring function:
\begin{equation}
    score = \sum_{cpl=0}^{255} s(cpl) \cdot cpl.
\end{equation}

We iterate through all CPLs in the solutions, multiplying the number of Sybil nodes in each CPL, obtained using the $s$ function, by their CPL. The function $s$ compares the initial distribution to the resulting one, identifying the number of additional malicious nodes inserted at each CPL. In case of a score tie, the tie-breaking criterion is the distribution with the lowest $D_{KL}$.

\begin{figure}[ht]
    \centering
    \includegraphics[width=\columnwidth]{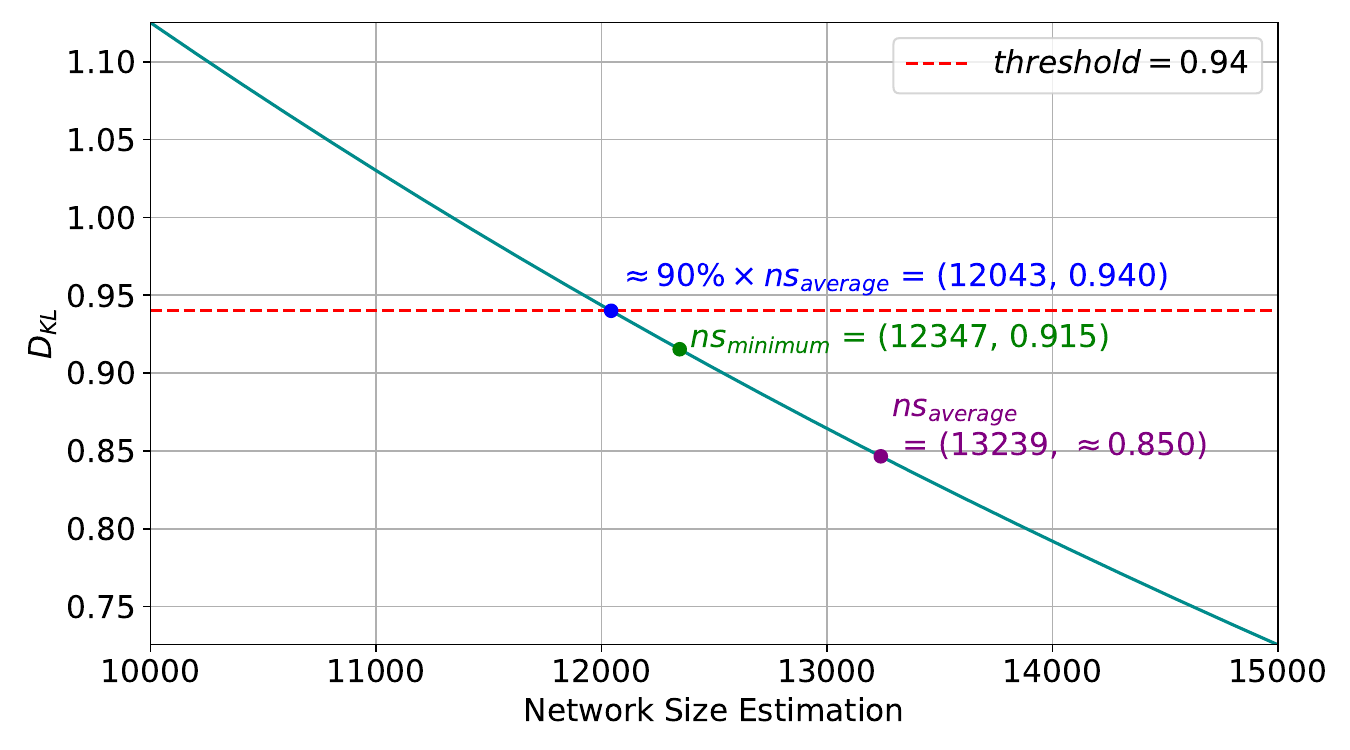}
    \caption{Impact of network size estimation on the calculation of $D_{KL}$ against an attacked distribution of $D_{KL} = 0.85$.}
    \label{sec4:fig:kl-comparison-from-ns}
\end{figure}

Since the network size estimation can vary between nodes, the model distribution and consequently the $D_{KL}$ value can also change. To avoid discrepancies between the attacker's and provider's estimations, we enforce a maximum $D_{KL} \leq 0.85$. This threshold accounts for an approximate 10\% margin of error in the network size estimation. When the provider underestimates the network size, the attack becomes easier to detect, as the model distribution tolerates fewer close nodes. To evaluate the impact of this variance on detection, we compared an attack generated with $D_{KL} \approx 0.85$ using an average estimation from a 24-hour test period, $ns_{average} = 13239$, against a detection performed using different network size estimations. When tested with the lowest recorded network size from that period $ns_{lowest} = 12347$, the attack remained undetected. These results are illustrated in Figure \ref{sec4:fig:kl-comparison-from-ns}.

\subsection{Cost Analysis}

The bandwidth and computing resources needed to instantiate the Sybil nodes are not considered significant, as a publicly accessible IP address and a stable internet connection are sufficient to carry out the attack. The primary cost of executing the active attack is the computational effort required to brute-force Sybil identifiers. While resource-intensive, this process can be completed in a relatively short time. 

\textbf{Brute-forcing Identifiers.} To obtain the malicious nodes, we brute-force identifiers by generating Ed25519 \citep{josefsson-edwards-curve-digital-signature-2017} key pairs. The expected number of attempts required to generate a single identifier within a specific CPL is given by $2^{\text{CPL} + 1}$. Consequently, the time required to generate a node at a given CPL can be expressed as:
\begin{equation}
    t(CPL) = t_{Ed25519} \cdot 2^{CPL + 1}.
\end{equation}

The constant $t_{\text{Ed25519}}$ represents the time required to generate an Ed25519 key pair using our implementation. On a machine equipped with a 13\textsuperscript{th} Gen Intel i7-13700H processor, $t_{\text{Ed25519}}$ is approximately $1.35\,\mu s$. Consequently, generating a node with $CPL = 25$ to the target CID would take approximately 1 minute and 30 seconds. While such CPL is considered exceptionally close to the target for a network of size $ns_{average}$, and was not observed in our tests, such proximity could still be brute-forced in a relatively short time.

\subsection{Experiment Description}

\begin{figure}[ht]
    \centering
    \includegraphics[width=\columnwidth]{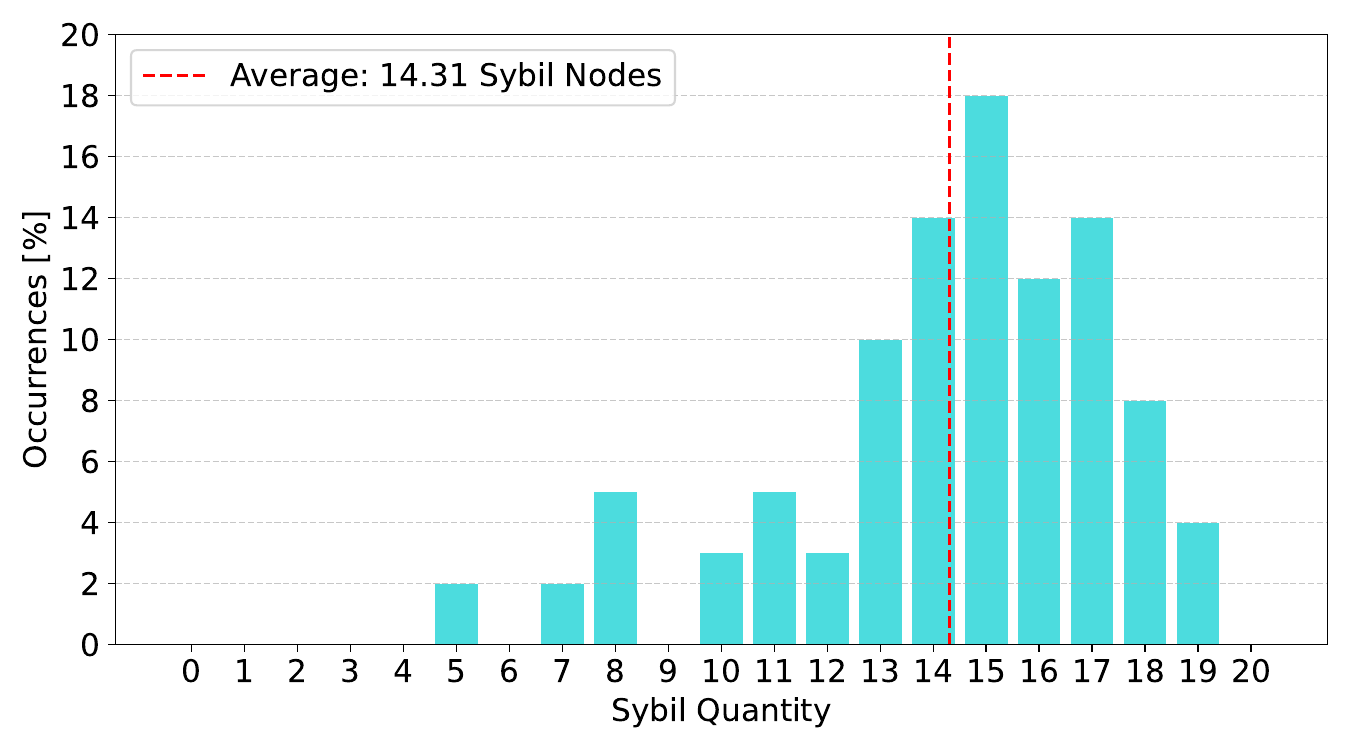}
    \caption{Optimized Sybil placement count across 100 random distributions.}
    \label{sec4:fig:sybil-optimization-quantity}
\end{figure}

Since the number of malicious nodes depends on the distribution of the $k$ closest peers, we first conducted an experiment to determine the average number of Sybil nodes that could be inserted into random peer distributions. As illustrated in Figure \ref{sec4:fig:sybil-optimization-quantity}, the experiment shows an average of $14.31$ Sybil nodes among the 20 closest. For this reason, to represent an average-case scenario, all our tests were conducted using distributions that allow the insertion of 14 malicious nodes.

In the same Figure \ref{sec4:fig:sybil-optimization-quantity}, we observe that in at least 91\% of the distributions, a minimum of 10 malicious nodes were successfully inserted. Throughout our tests, we did not encounter any distribution in which inserting Sybil nodes was impossible despite the detection algorithm false-positive rate (see Figure \ref{sec4:fig:kl-optimization-distribution}). 
 This is due to our optimization algorithm correcting de facto distributions that would have led to a false positive case by transforming the set of $k$ closest nodes into a valid distribution that remains below the detection threshold.

To ensure the reliability of our results, five attacks were executed in parallel against five distinct distributions, involving a total of $14 \times 5 = 70$ Sybil nodes operating simultaneously. 
Each Sybil node ran a modified Kubo instance and was deployed within a public network, ensuring their accessibility for DHT requests, responding to queries for the targeted content with ten randomly generated provider records. These nodes were interconnected, enabling them to recommend one another during lookup operations and ensure the consistent capture of 14 out of the 20 available records in each content publication.

After the Sybil nodes were instantiated, the CID was announced by a node running both the attack detection and the region-based mitigation. Upon detecting an attack, this node would automatically distribute the provider records across the entire mitigation region. The requester, which verifies whether the content had been eclipsed, used unmodified Kubo implementations, since the defense mechanism only protects the publication process. These nodes searched for the content every 30 minutes and were restarted every 60 minutes. Content was considered unreachable if no response was retrieved within 30 seconds.

In an attack scenario targeting existing content in the network, the corresponding provider records would have been published prior to the instantiation of Sybil nodes. However, in our experiment, we opted to instantiate the malicious nodes before the publication phase to avoid waiting for the initial expiration cycle, which would otherwise delay the results by 48 hours. While the results would remain similar, the effectiveness of the attack would be reduced during the first two days.

\subsection{Environment} \label{sec4:ssec:environment}

Three types of nodes are necessary for our experiments: \textit{(i)} Sybil nodes; \textit{(ii)} a provider implementing both the attack detection and region-based mitigation mechanisms; and \textit{(iii)} requesters.  Each node has its own public IP address to ensure unrestricted accessibility through DHT queries. For \textit{(i)}, all Sybil nodes were executed simultaneously on a single machine equipped with an Intel i7-9700 CPU @ 3.00 GHz and 32 GB of RAM. For \textit{(ii)}, the provider was deployed on a separate computer and network, in an OVH cloud virtual machine equipped with one Intel Core Processor (Haswell, no TSX), two cores, and 4 GB of RAM. Finally, the requesters were also deployed on a different network, running on a machine equipped with an Intel(R) Xeon(R) CPU E5-2640 v4 @ 2.40 GHz, with 10 cores and 110 GB of RAM, which is more powerful than necessary for the performed tasks.

As further discussed in Section \ref{ec} on \emph{Ethics}, only our own random files were eclipsed during our experiments. This ensures that no denial of service is caused for content actually being shared on the network. Additionally, the nodes inserted into the network remained collaborative for any other action but related to the targeted CIDs, thereby fully contributing to the network.

\subsection{Evaluation}
\label{sec4:ssec:active-attack-evaluation}

\begin{figure}[ht]
    \centering
    \includegraphics[width=\columnwidth]{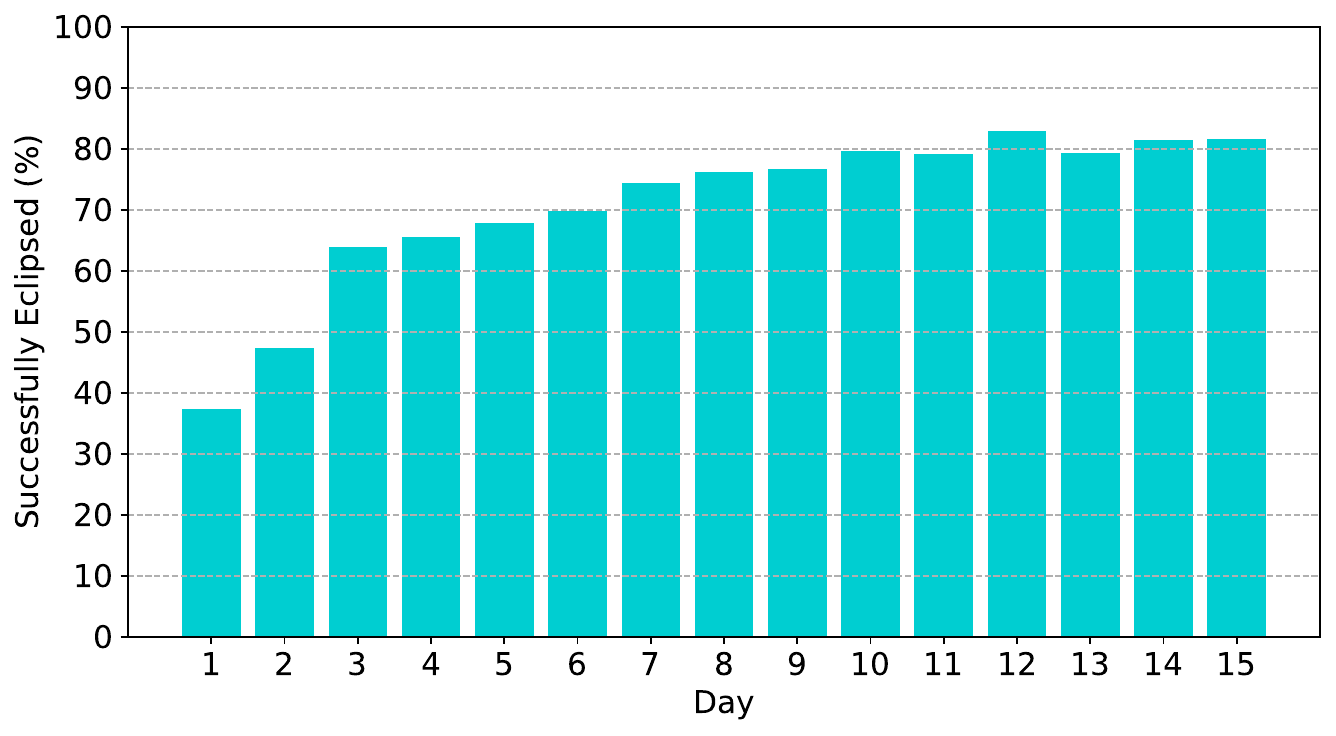}
    \caption{Active attack using optimized Sybil placement against region-based mitigation.}
    \label{sec4:fig:active-attack-evaluation-barplot}
\end{figure}

The results of our attack are illustrated in Figure \ref{sec4:fig:active-attack-evaluation-barplot}, which shows the eclipse percentage during content retrieval over a 15-day period. By the end of the second day, the eclipse rate had already reached nearly 50\%. This rate continued to increase over time, due to the natural adaptation period required for the Sybil nodes. Since routing tables in Kademlia favor stable and long-standing peers, malicious nodes must wait for opportunities to be gradually incorporated into more routing tables.

Starting from the 10\textsuperscript{th} day, the results began to stabilize, converging to an average eclipse rate of 80\%. By the end of the experiment, this rate had slightly increased to 82\%. Although the test was extended for an additional 15 days, the results remained consistent and are therefore not included in the figure.

\subsection{Discussion}

While Sybil attacks are an inherent problem of the DHT design, the effectiveness of our optimized active Sybil attack is improved by a Kubo implementation choice, which halts the lookup procedure after receiving only ten provider records from a single source. Although this behavior is specific to Kubo, other IPFS implementations or Kademlia-based networks may also be vulnerable to this attack strategy, since the number of provider records accepted before halting must remain bounded to end a successful search while avoiding introducing new attack vectors (such as flooding and message overflow).

%% file: src/section/5/section-5.tex
This section introduces a new defense mechanism for mitigating Sybil attacks in DHTs: the Sybil-Resistant DHT Store (SR-DHT-Store). The proposed approach is designed to be executed during every publication, securing the provide operation while imposing minimal overhead. We begin by describing the defense mechanism, followed by an analysis of its internal structure, and conclude with an evaluation of its performance and associated costs. For evaluation, the mitigation was fully implemented in the Kademlia libp2p repository. 

\subsection{Description}

PIDs should be randomly assigned and uniformly distributed across the Kademlia address space. In a network of $N$ nodes,  the expected distance $d_k$ at which the closest $k$ nodes should be found is:
\begin{equation}
    d_k = \frac{\text{keySpace}}{N} \times k.
\end{equation}

However, estimating $N$ in a decentralized environment, where nodes lack a global view of the network, is challenging and introduces significant overhead. To avoid computing the network size, we leverage the structure of the routing tables and query known nodes to obtain a sufficiently accurate estimation of the distance $d_k$.

\begin{figure}[ht]
    \centering
    \includegraphics[width=\columnwidth]{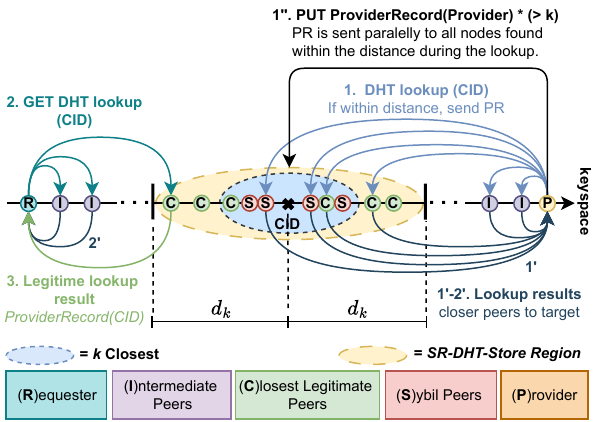}
    \caption{SR-DHT-Store opportunistic publication. This example assumes $k = 6$.}
    \label{sec5:fig:esr-store}
\end{figure}

When performing a lookup to identify the $k$ closest nodes during publication, and given that $d_k$ has been previously calculated, we can opportunistically send provider records to any nodes found that are closer than the calculated distance during the lookup. This exchange can be executed in parallel while continuing the lookup process. Just after completing the lookup, if the total number of sent records is below the replication factor $k$, additional records are distributed to nodes beyond $d_k$ until $k$ peers store the data. In the presence of an attack, both reliable and Sybil nodes will receive the records, causing the total number of resolvers to exceed $k$ by approximately the number of instantiated Sybil nodes. The mitigation process is illustrated in Figure \ref{sec5:fig:esr-store}.

To summarize, our approach differs from Kademlia \citep{maymounkov-kademlia-2007} in that we no longer rely on the closest nodes to a CID, which favors Sybil Attacks but by defining a proper distance. We diverge from the work of \citet{Sridhar et al.}{sridhar-content-censorship-2023} as we do not seek to detect the attack, since the detection can be cheated, but rather to improve the way region-based queries can be systematically performed, as explained in the following sections. 

Additionally, our distance estimation employs the XOR metric instead of the CPL used in \citet{Sridhar et al.}{sridhar-content-censorship-2023}. Unlike the CPL, the XOR metric is free from threshold effects, as CPL-based zones are limited by powers of two, only doubling or halving based on the estimated network size. An inaccurate estimation in the CPL, can lead to additional nodes being contacted to store records in case of detection. As of August 2024, the average estimated CPL is around 9, typically enclosing $k = 20$ nodes. However, with network churn, the region boundary may fall between CPL values, leading to incorrectly defined regions and significant variation in the number of nodes receiving the record.

\subsection{Zone Delimitation}

While an accurate estimation of the average $d_k$ value is crucial for our mitigation, this calculation must also be efficient to minimize overhead. We identified two methods for determining the distance to the $k$-th farthest node from random peers in order to calculate the average $d_k$: (i) directly requesting their $k$ closest nodes, or (ii) performing a lookup toward random IDs. Due to the structure of the routing table, which is organized into $k$-sized buckets, nodes store more information about their closest peers than about the rest of the network. Our approach takes advantage of this by directly querying nodes for their $k$ closest peers. While this method is not as precise as a lookup, it provides a sufficiently accurate estimation with minimal overhead.

To evaluate this trade-off, we compared the $k$ closest nodes obtained from both approaches. On an average of 100 tests, $16.32$ among the $k = 20$ closest nodes were identical in both methods. Despite this minor difference, each lookup involved querying an average of 42.3 nodes until reaching termination, while a direct query required only a single request. For this reason, when joining the network, we opted to obtain the initial average $d_k$ value by querying $Q_{d_k}$ peers for the closest nodes they know and calculating the average distance of the $k$-farthest node.

\begin{figure}[ht]
    \centering
    \includegraphics[width=\columnwidth]{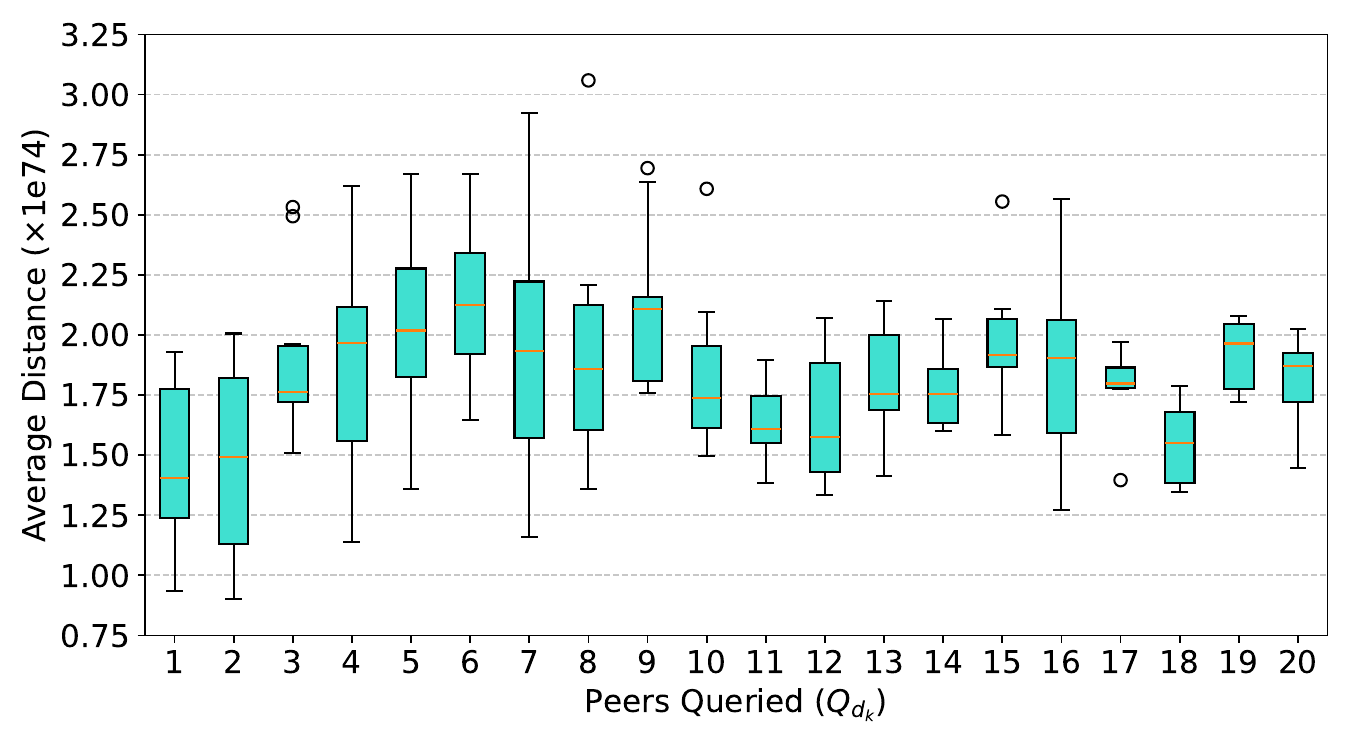}
    \caption{$Q_{d_k}$ variation impact on the average distance of the $k$-th closest node.}
    \label{sec5:fig:peer-queried-comparison-bar.pdf}
\end{figure}

In Figure \ref{sec5:fig:peer-queried-comparison-bar.pdf}, we evaluate the impact of querying $1 \leq Q_{d_k} \leq 20$ random peers for the calculation of $d_k$. 
Each value of $Q_{d_k}$ within the range was evaluated ten times, and the boxes are computed based on the average distances from those 10 measurements. As observed in the chart, measurements with $Q_{d_k} < 10$ exhibit a greater span that could often lead to a less precise initial estimation of $d_k$ (except for $Q_{d_k} = 16$, which seems to be an outlier). 
In fact, local fluctuations inevitably appear around the average because each distance measurement can be considered as a sample of a distribution function, but also because the DHT lookup may sometimes fail to accurately find all the exact K-closest peers.
For this reason, during initialization, the node first queries $Q_{d_k} = 10$ peers to compute an initial average $d_k$, which provides a sufficiently accurate distance to begin the provide procedure.

\subsection{Churn-Aware Distance Estimation}

To account for network size fluctuations and improve the accuracy of the estimation, the average distance to the $k$-farthest peer ($d_k$) must be updated over time. Typically, P2P networks are subject to hourly size variation which magnitude depends on the proportion of personal computers hosting peers using public IP addresses. To maintain an accurate average $d_k$, each time a lookup is performed, the distance of the $k$-farthest found node should be incorporated into the $d_k$ calculation. 

As explained in \ref{sec2:ssec:routing-table}, the routing table is updated every ten minutes through $L_{d_k} = 16$ lookups, which can be used to refine our average distance estimation with no additional overhead. To integrate new measurements into $d_k$ while preserving the influence of previous values, we propose using an Exponentially Weighted Moving Average (EWMA). This type of average allows updates while maintaining stability in the estimation. The updated distance can be calculated using the following equation:
\begin{equation}
    S_{t+1} = y_t \cdot \alpha_{sf} + S_t (1 - \alpha_{sf}).
\end{equation}

In the equation, $y_t$ denotes the latest measurement, $S_t$ the previously calculated moving average, and $\alpha_{sf} \in [0, 1]$ the smoothing factor, which determines the influence of past measurements on the average. According to \citet{Hunter}{the-ewma-hunter-1986}, the optimal value of $\alpha_{sf}$ minimizes the mean squared error (MSE) relative to the target - here, the distance that returns exactly $k$ provider records. Hunter also recommends initializing with a simple average over 4–5 values, which we follow by initially estimating $d_k$ using $Q_{d_k} = 10$ queries. To determine the optimal $\alpha_{sf}$ for our scenario, we first estimated $d_k$ with $Q_{d_k} = 10$, then refined it using $L_{d_k} = 16$ lookups, evaluating the MSE across $\alpha_{sf}$ values ranging from $0.001$ to $0.999$ in increments of $0.001$. The lowest average MSE was approximately at $\alpha_{sf} = 0.1$.

For the rest of the paper, we estimate $d_k$ by measuring the distance to the $k$-th farthest node from the results of the $k$ closest nodes from  $Q_{d_k} = 10$ queries and $L_{d_k} = 16$ additional lookups. The moving average is updated with a weighting factor of $\alpha_{sf} = 0.1$. By using this strategy, we take into account both the initial distance estimation and the subsequent lookups performed during the node's initialization to populate its routing table.

\subsection{Cost Analysis}

\begin{figure}[ht]
    \centering
    \includegraphics[width=\columnwidth]{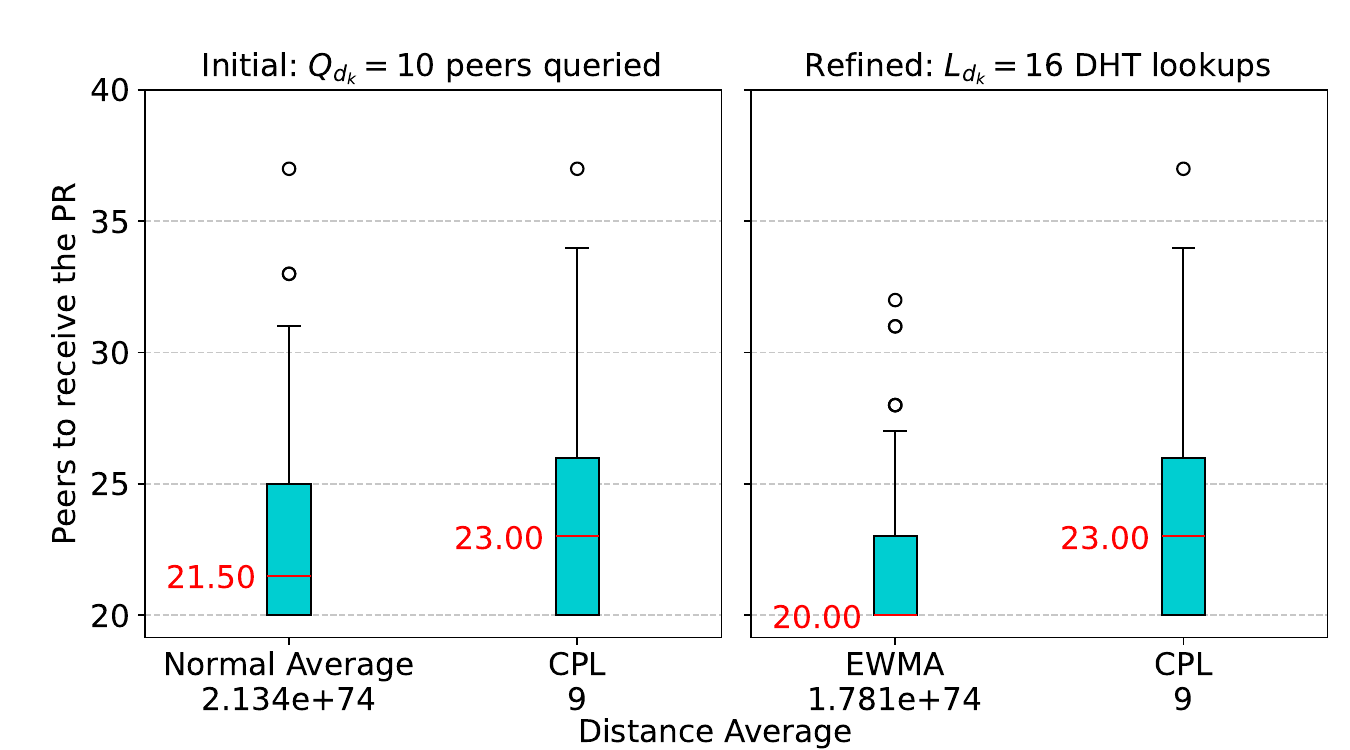}
    \caption{Average number of nodes receiving PRs when initializing $d_k$ with $Q_{d_k} = 10$ queries and refining with $L_{d_k} = 16$ lookups.}
    \label{sec5:fig:esr-store-peers}
\end{figure}

To evaluate the cost of our mitigation strategy, we assessed the accuracy of our $d_k$ estimation using a dataset containing all peers obtained from 100 random DHT lookups. The goal was to determine how many additional nodes (overshoot) would have received the provider records based on the estimated distance. The distance calculation was estimated and tested ten times against the dataset, and the results are presented in Figure \ref{sec5:fig:esr-store-peers}. 

The Figure \ref{sec5:fig:esr-store-peers} is divided into two parts: one showing the initial distance estimation (based on simple queries) and the other presenting the refined estimation using EWMA (derived from lookup results). The x-axis shows the average distance calculated across the ten tests for both the initial and refined phases. All data points from each of the ten tests were included in the chart, meaning that each box plot represents $10 \text{ (tests)} \times 100 \text{ (lookups per test)} = 1000$ data points. Each chart also includes the equivalent CPL-based approach, which corresponds to sending provider records to all nodes within the CPL range defined by $keyspace - \lfloor\log_2(d_k)\rfloor$, similarly to the region-based queries mitigation.

By querying only $Q_{d_k}$ nodes for their $k$-th farthest peer, we observed a median of $k + 1.5$ nodes receiving the provider record. After refinement, the median matched exactly $k$. In distributions where fewer than $k$ nodes were located within the distance $d_k$, the node continues the publication process until reaching the target of $k$ peers, selecting slightly more distant nodes as necessary. To evaluate the cost associated with this procedure, we consider only the additional provider record exchanges that exceed $k$, since the transmission of $k$ records is already part of the standard, unprotected provide operation.

Before and after refinement, the average number of additional nodes contacted was respectively $C_{XOR_i} = 2.786$ and $C_{XOR_r} = 2.064$, respectively. When using CPL-based zones to define the target region instead, the number of contacted nodes increases to $C_{CPL_i} = 5.135$ and $C_{CPL_r} = 3.860$. For a newly instantiated node, the total overhead introduced by our mitigation is $C = Q_{d_k} + C_{XOR_i} = 12.786$ contacted nodes. For already established nodes, regular lookups naturally refine the precision of $d_k$, reducing the extra cost to $C_{XOR_r} = 2.064$ contacted nodes per publication.

\textbf{Cost Comparison.} The number of nodes contacted at each stage of the provide procedure in the default Kubo implementation, the attack detection with region-based mitigation, and the SR-DHT-Store is compared in Table \ref{sec5:tab:cost-comparison}.
The table highlights mitigation-specific phases, and shared phases, such as the main lookup performed to find the $k$ closest nodes to the content.

{
    \definecolor{phase}{HTML}{FFD966}
    \definecolor{phase-light}{HTML}{FFF2CC}
    \definecolor{standard-provide}{HTML}{9ABA84}
    \definecolor{region-based-queries}{HTML}{D5A6BD}
    \definecolor{sr-dht-store}{HTML}{F6B26B}
    \definecolor{attack}{HTML}{9FC5E8}
    \definecolor{node-type}{HTML}{68CBD0}
    \tiny

    \begin{table}[ht]
        \centering
        \caption{Number of contacted nodes during each phase of the provide procedure in the default Kubo implementation, attack detection + region-based queries and  SR-DHT-Store.}
        \label{sec5:tab:cost-comparison}
        \begin{tblr}{width=\columnwidth, colspec={|X[0.2,c]|X[0.2,c]|c|c|c|}}
            \hline
            \SetCell[c=2, r=2]{c, phase} Phase && \SetCell[c=3]{c, phase} \makecell{Number of Contacted Nodes} && \\
            \hline
            && \SetCell{standard-provide} \makecell{Default Kubo} & \SetCell{region-based-queries} \makecell{Attack Detection + \\ Region-Based \\ Queries \citep{sridhar-content-censorship-2023}} & \SetCell{sr-dht-store} SR-DHT-Store \\
            \hline
            \SetCell[c=2]{c, phase-light} \makecell{Network size \\ Estimation \\ after Bootstrap} && - & 433.8 & - \\
            \hline
            \SetCell[c=2]{c, phase-light} \makecell{$d_k$ Estimation} && - & - & 10 \\
            \hline
            \SetCell[c=2]{c, phase-light} \makecell{Lookup} && \SetCell[c=3]{c} 42.3 && \\
            \hline
            \SetCell[c=2]{c, phase-light} \makecell{Region-Based \\ Queries} && - & 72.9 & - \\
            \hline
            \SetCell[c=2]{c, phase-light} \makecell{False Positives} && - & $72.9 \times 0.11 \approx 8$ & - \\
            \hline
            \SetCell[c=2]{c, phase-light} \makecell{PR Overshoot} && - & - & 2.7 to 2 \\
            \hline
        \end{tblr}
    \end{table}
}

Immediately after node initialization, their attack detection mechanism requires estimating the network size. According to their implementation on Kubo\footnote{\url{https://zenodo.org/records/8300034}}, this procedure involves performing 10 lookups to gather network data, which, based on our experiments, results in contacting an average of 433.8 nodes. We avoid this overhead by directly estimating the distance to the $k$-th closest peer, thereby eliminating the need to infer the network size in order to derive $d_k$, as this value is already available from the lookup responses. This approach not only reduces unnecessary overhead but also avoids the inaccuracies introduced when estimating the publication zone based on network size.

In our approach, the full list of peers discovered during the main lookup process, obtained when looking for the $k$ closest nodes to the content during publication, is kept and reused, eliminating the need for a separate region crawl. Indeed, while Kademlia typically returns only the $k$ closest nodes from a lookup, it is possible to retain all discovered peers encountered during the process. 
To compare the peer lists obtained through region-based queries and standard lookups, we identified all nodes sharing at least a CPL of nine over ten randomly selected identifiers. On average, the dedicated region-based queries discovered 22.9 nodes, while standard lookups returned 22.7 nodes. Since the standard lookup is already part of the standard publication procedure and obtains similar results when compared to the region crawl, we consider the entire cost of region-based queries as additional overhead.

In conclusion, Table \ref{sec5:tab:sum-cost-comparison} shows the total number of contacted nodes obtained by summing each individual phase of the provide procedure.
For newly instantiated nodes, considering the 11\% false positive rate in attack detection (as shown in Figure \ref{sec4:fig:kl-optimization-distribution}) and our initial provider record overshoot of 2.7 nodes, SR-DHT-Store reduces the number of contacted nodes by approximately 88.6\% in the absence of attacks and nearly 90\% during attacks. For nodes already established in the network, the overhead is reduced by 10\% under normal conditions and by 63\% in the presence of an attack. Compared to an unprotected client, the overhead of SR-DHT-Store is only 4\% for an established peer. 

{
    \definecolor{phase}{HTML}{FFD966}
    \definecolor{phase-light}{HTML}{FFF2CC}
    \definecolor{standard-provide}{HTML}{9ABA84}
    \definecolor{region-based-queries}{HTML}{D5A6BD}
    \definecolor{sr-dht-store}{HTML}{F6B26B}
    \definecolor{attack}{HTML}{9FC5E8}
    \definecolor{node-type}{HTML}{68CBD0}
    \tiny

    \begin{table}[ht]
        \centering
        \caption{Total number of contacted nodes obtained by summing each phase of the provide procedure for the default Kubo implementation, the attack detection with region-based queries, and SR-DHT-Store.}
        \label{sec5:tab:sum-cost-comparison}
        \begin{tblr}{width=\columnwidth, colspec={|X[0.20,c]|X[0.14,c]|X[0.16,c]|X[0.3,c]|X[0.2,c]|}}
            \hline
            \SetCell[r=2]{c, phase} \makecell{Node DHT \\ Status} & \SetCell[r=2]{c, phase} \makecell{Attack \\ Scenario} & \SetCell[c=3]{c, phase} \makecell{Total Number of Contacted Nodes} && \\
            \hline
            && \SetCell{standard-provide} \makecell{Default \\ Kubo} & \SetCell{region-based-queries} \makecell{Attack Detection \\ + \\ Region-Based \\ Queries \citep{sridhar-content-censorship-2023}} & \SetCell{sr-dht-store} SR-DHT-Store \\
            \hline
            \SetCell[r=2]{c, node-type} \makecell{Newly \\ Instantiated} & \SetCell{c, attack} No & \SetCell{c} 42.3 & \SetCell{c} 484.1 & \SetCell{c} \textbf{55} \\ 
            \hline
            & \SetCell{c, attack} Yes & \SetCell{c} 42.3 & \SetCell{c} 549 & \SetCell{c} \textbf{55} \\ 
            \hline
            \SetCell[r=2]{c, node-type} \makecell{Already \\ Established} & \SetCell{c, attack} No & \SetCell{c} 42.3 & \SetCell{c} 50.3 & \SetCell{c} \textbf{44.3} \\ 
            \hline
            & \SetCell{c, attack} Yes & \SetCell{c} 42.3 & \SetCell{c} 115.2 & \SetCell{c} \textbf{44.3} \\ 
            \hline
        \end{tblr}
    \end{table}
}

\subsection{Evaluation}
\label{sec6:evaluation-of-sr-dht-store}

The SR-DHT-Store was implemented into Kademlia libp2p, and evaluated against both the passive $k$ closest attack and the active optimized Sybil placement attack, illustrated on Table \ref{sec6:tab:sr-dht-store-against-mitigations}. All Sybil nodes shared the same implementation, with configurable flags to toggle their behavior, enabling them to switch between passive and active modes. 
The environment was configured in the same way as described in Section \ref{sec4:ssec:environment}, except that the provider was using the SR-DHT-Store.

{
    \definecolor{retrieval-success-rate}{HTML}{68CBD0}
    \definecolor{attack}{HTML}{FFD966}
    \definecolor{attack-light}{HTML}{FFF2CC}
    \definecolor{region-based-queries}{HTML}{D5A6BD}
    \definecolor{sr-dht-store}{HTML}{F6B26B}

    \begin{table}[ht]
        \centering
        \caption{Evaluation of both mitigation strategies against the passive $k$ closest and the active optimized Sybil positioning attacks.}
        \label{sec6:tab:sr-dht-store-against-mitigations}
        \begin{tblr}{width=\columnwidth, colspec={|X[0.30,c]|X[0.36,c]|X[0.33,c]|}}
            \hline
            \SetCell[r=2]{attack} Attack & \SetCell[c=2]{ retrieval-success-rate} Retrieval Success Rate (\%) & \\ 
            \hline
            & \SetCell{region-based-queries} \makecell{Attack Detection + \\ Region-Based Queries} & \SetCell{sr-dht-store} SR-DHT-Store \\
            \hline
            \SetCell{attack-light} Passive $k$ Closest & 100\% & 100\% \\
            \hline
            \SetCell{attack-light} \makecell{Active Optimized \\ Sybil Positioning} & 18\% & 28\% \\
            \hline
        \end{tblr}
    \end{table}
}

We targeted five distinct CIDs, publishing the content using both mitigation strategies: SR-DHT-Store and the combination of attack detection with region-based queries. Following publication, we conducted standard lookups to retrieve the provider records by instantiating a peer in a completely separate network. As a result, the content was successfully retrieved in every search, achieving a 100\% success rate for both mitigation approaches.

For the optimized Sybil placement attack, we reused the same Sybil nodes deployed in our earlier active attack experiment (Section \ref{sec4:ssec:active-attack-evaluation}) to evaluate the mitigations against well-established adversarial nodes. Since both mitigation strategies were tested against the same group of Sybil nodes, each CID was published from different peers depending on the mitigation used. During retrieval, if the test-specific provider was missing from the set of providers listed in the PRs, we considered the content to be eclipsed, as the response was assumed to originate from a Sybil node. Similarly to the passive attack scenario, each test was repeated five times for each mitigation strategy. As a result, the region-based query approach, which initially failed to detect the attack, worked as a standard lookup mechanism, retrieving the right provider in only 18\% of cases. When using SR-DHT-Store, we increase 10\%, leading to 28\% of the lookups successfully retrieving the content. It sounds disappointing but it can be improved. 

The rationale behind the low success rate is as follows: since some active Sybil nodes are among the closest to the content, they may still be the first contacted during content retrieval. This can lead to an early abortion of the lookup process, even when other nodes in the network hold the correct provider record. Our mitigation focuses on securing the provide procedure to ensure that provider records are not exclusively received by malicious nodes. However, the search process remains vulnerable to active attacks due to Kubo's termination behavior, and thus requires further improvements to address the threat. In the following section, we introduce additional mitigation strategies and defense mechanisms that can complement our approach, along with their corresponding evaluation.

%% file: src/section/6/section-6.tex
While SR-DHT-Store secures the content publication phase, the lookup procedure remains vulnerable to adversaries exploiting early termination behavior. To address this, we propose in this section a combination of SR-DHT-Store with two client-side defense mechanisms, both implemented in Kubo and evaluated against our active attack model. An additional mechanism is also introduced, however it was not implemented, as we believe it increases the difficulty of performing the attack but does not fully mitigate the vulnerability.

\textbf{PR Limitation.} No external node should be able to prematurely terminate a DHT lookup alone. Currently, the maximum number of provider records required to stop a lookup, $PR_{\text{max}}$, and the maximum number of provider records accepted from a single peer, $PR_{\text{max}_{peer}}$, are both set to ten, i.e., $PR_{\text{max}} = PR_{\text{max}_{peer}} = 10$. This configuration allows a single external peer to return just enough records to externally halt the whole lookup process. 

To prevent this behavior, setting $PR_{\text{max}} > PR_{\text{max}_{peer}}$ ensures that the node continues searching until it has locally accumulated $PR_{\text{max}}$ provider records. This approach preserves compatibility with other peers by leaving $PR_{\text{max}_{peer}}$ unchanged, and only the local threshold $PR_{\text{max}}$ is adjusted to decide when to terminate the lookup based on the total number of collected records.

In our implementation, we tested two different values for $PR_{\text{max}}$: 50 and 200. These values correspond to requiring complete sets of ten provider records from 5 and 20 peers, respectively, before externally halting the lookup.

\textbf{Disjoint Lookup Paths.} Proposed by \citet{Baumgart and Mies}{baumgart-kademlia-practicable-approach-2007}, this secure lookup mechanism was introduced in Section \ref{sec3:subsec:ipfs-defense-mechanisms}. In a standard Kademlia lookup, $\alpha$ nodes are queried in parallel for closer peers, but all discovered nodes are added to a shared query list. This design allows a single adversarial node, once queried, to compromise the entire search. To mitigate this vulnerability, we employ $d = 3$ disjoint lookups, in which the results of each path are kept entirely separate. Additionally, no node should be contacted more than once across the different paths. Although this approach introduces additional overhead, it could be used selectively, only when searching for content, rather than for all lookup operations. To further minimize the overhead, disjoint lookups can be triggered only when the initial lookup attempt fails to retrieve the desired content.

\textbf{IP Address Limitation in Lookup.} During a DHT lookup, no more than one node per IP address, or per a given IP subnet prefix, should be contacted. This limitation increases the difficulty for attackers, as it requires a larger number of unique IP addresses to successfully carry out an attack. When combined with disjoint lookup paths and the PR limitation, it introduces an additional layer of complexity, forcing the attacker to distribute Sybil nodes across multiple networks. While this measure increases the effort required to launch an attack, it is not a definitive solution as the other proposed mitigations. For this reason, it was neither implemented nor evaluated in our experiments.

\subsection{Cost Analysis}

The PR limitation introduces no additional cost compared to a standard lookup when no attack is present. In the absence of an attack, most nodes will typically return valid provider records for the content, and before reaching $PR_{\text{max}}$, the providers will eventually have the content. In the case of an active attack, however, each contacted Sybil node may trigger up to ten additional peer searches for the fake PIDs. This occurs because the node is initially unable to contact the providers returned by the malicious peers, due to invalid or unreachable IP addresses, and must perform a separate lookup for each false provider in an attempt to resolve valid contact information.

The second mitigation, disjoint requests, introduces an overhead of approximately 2.5 times the cost of a standard lookup, as it performs $d = 3$ independent searches. To quantify this overhead, we conducted ten lookups using both the standard and disjoint request approaches, querying for previously published content. On average, the standard lookup contacted 21.9 nodes before retrieving a valid provider record, whereas the disjoint requests contacted an average of 55.7 nodes.

\subsection{Evaluation}

\begin{figure}[ht]
    \centering
    \includegraphics[width=\columnwidth]{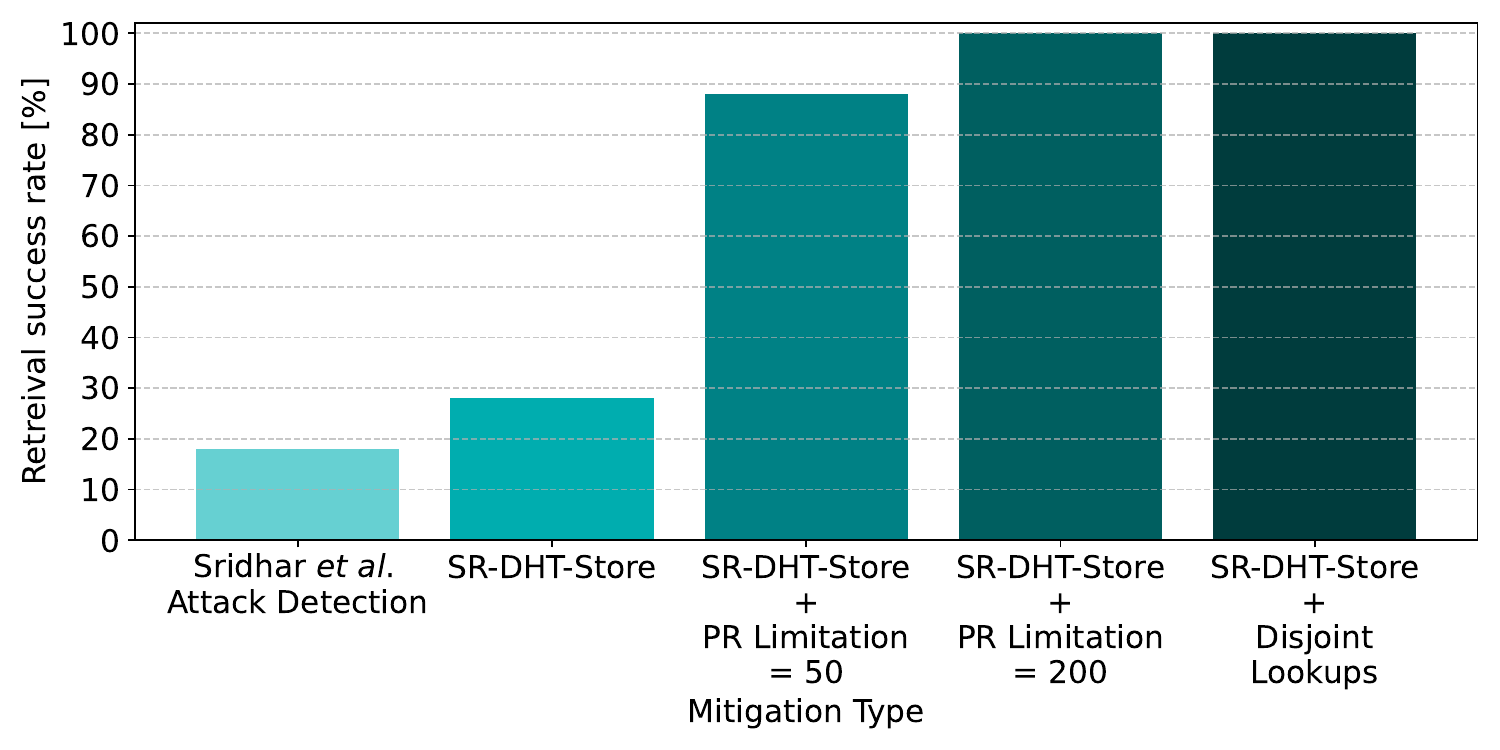}
    \caption{Effectiveness of mitigations against the optimized Sybil positioning attack.}
    \label{sec7:fig:enhanced-mitigation}
\end{figure}

All tests were conducted against the same Sybil nodes previously instantiated for the evaluation of our active Sybil attack, described in Section \ref{sec4:ssec:active-attack-evaluation}. We began by evaluating only the publication-phase mitigations, followed by a standard lookup for content retrieval. Afterward, we combined the SR-DHT-Store publication mechanism with lookups applying both the PR limitation, using thresholds of 50 and 200, and disjoint lookup paths. Each configuration was tested across five distinct CIDs, with five retrieval attempts per CID. The results are presented in Figure \ref{sec7:fig:enhanced-mitigation}.

Similarly to the active attack and the standalone SR-DHT-Store, the environment was configured as described in Section \ref{sec4:ssec:environment}, except that the provider used the SR-DHT-Store and the requesters implemented the client-side mitigations.

The first two columns in Figure \ref{sec7:fig:enhanced-mitigation} replicate the results from Section \ref{sec6:evaluation-of-sr-dht-store}. The last three columns present the success rate when combining SR-DHT-Store with retrieval-phase defenses. With a PR limitation of 50, content was successfully retrieved in 88\% of cases, demonstrating improved resistance. However, this also reveals the persistence of the Sybil nodes: in 12\% of the lookups, five malicious peers were among the first contacted. Increasing the threshold to 200 or employing disjoint lookup paths resulted in a 100\% success rate, fully mitigating the attack.

\subsection{Discussion}

We should note that the lookup mitigations considered in this section depend on the presence of a publication-side defense, such as the SR-DHT-Store. While these mechanisms may provide some resistance against the optimized Sybil placement attack, they are ineffective in scenarios where all $k$ closest nodes are malicious, as all provider records are then stored on untrustworthy nodes, rendering content irretrievable. For this reason, the retrieval-side mitigations are not evaluated alone, as they remain vulnerable to even basic attack strategies.

When combined with the SR-DHT-Store, the PR Limitation works well against active attack strategies, while disjoint lookups can be applied against both active and passive attacks, but also against adversarial routing \citep{kohnen-conducting-optimizing-eclipse-2009}. We consider the PR Limitation a simple and efficient solution that could be implemented to counter active attacks, at almost no cost. The disjoint lookups offer a more robust and general solution that brings diversity in the result of a DHT search to better withstand Sybil attacks. While disjoint requests increase the cost of performing lookups, they are capable of fully mitigating Sybil attacks on DHTs when applied after the SR-DHT-Store. In conclusion, we highlighted synergies between our provide mitigation and two other lookup defense mechanisms that leave some choice for IPFS developers to select the defense strategy based on the acceptable overhead.

Finally, we would like to highlight the fact that, while tested on IPFS, the mitigation strategy we propose is fully compatible with the standard Kademlia design and can be directly applied to protect any other P2P network running a DHT using the XOR metric (BitTorrent's Mainline DHT, KAD, Ethereum Swarm, etc.).

%% file: src/conclusion.tex
In this paper, we presented a new active Sybil attack targeting Kubo, the largest IPFS implementation. The attack was evaluated against the latest mitigation proposed by Sridhar et al. \citep{sridhar-content-censorship-2023}, which, although well implemented in a forked version, has not yet been merged into the official repository. Our attack achieved a success rate of 82\% against their mitigation by exploiting an early termination in the Kubo lookup process and an optimized placement of Sybil nodes to stay below the detection threshold triggering the mitigation. When executed against the current official version of Kubo, the attack is capable of completely eclipsing content, similar to the previously identified passive Sybil attack on the network.

To mitigate the attack, we introduced the SR-DHT-Store, a low-cost defense mechanism executed during each content publication. While our approach effectively addresses previously identified passive attacks on the network at a lower overhead, it does not, on its own, fully mitigate the active attack presented in this work. To achieve complete protection, we introduced two additional defense mechanisms designed to secure the content requester. When combined with our new publication strategy, both mechanisms fully mitigated the active attack.

In future work, we plan to explore new active attacker models that aim to disrupt the network not regarding the DHT’s behavior, but rather hijacking what is stored in it. Our goal is to counter content pollution by designing new consensus mechanisms that rely on the responses diversity enabled by SR-DHT-Store. Additionally, we plan to explore the efficiency and feasibility of our attack approach in other IPFS implementations and in Kademlia-based networks. From the performance perspective, we will evaluate 
the retrieval times of both the provide and lookup procedures with and without SR-SHT-Store and try to optimize the performance of DHT operations. 

%% file: src/ethics-consideration.tex
All targeted content was provided exclusively by nodes under our control. No legitimate content in the network was disrupted or eclipsed, since our malicious nodes responded normally to all requests except those specifically targeting the test content. To prevent instability in the DHT, the requester nodes operated in client mode, minimizing the impact of their joining and leaving the network.

While this paper introduces a new active attack approach targeting IPFS, the recently discovered passive attack remains unmitigated in the current Kubo implementation. Therefore, this experiment does not introduce any new threats beyond those already present in the system. Since the mitigation proposed by Sridhar et al. \citep{sridhar-content-censorship-2023} has not yet been integrated into the mainstream client, we do not consider this to be an attack on Kubo itself.

We contacted Protocol Labs, the organization behind IPFS, twice as part of our responsible disclosure process. The first, in 2023, served to inform them of our experiments on the network. The second, in 2025, aimed to disclose the newly discovered attack and to avoid them implementing a mitigation strategy that we found to be vulnerable.

%% file: src/acknowledgment.tex
This work was partly supported by the France 2030 ANR Project ANR-PECL-0009 TRUSTINCloudS and by the Inria Alvearium challenge.